\documentclass[a4paper,11pt]{article}

\pdfoutput=1
\usepackage{head,fullpage}     % Add local fullpage and head macros
\usepackage[pdftex]{graphicx}  % Add graphicx package with pdf flag (must use pdflatex)
\usepackage[utf8]{inputenc}
\usepackage{ulem}

\usepackage{hyperref}
\def\equationautorefname~#1\null{Eq.~(#1)\null}
\def\figureautorefname~#1\null{Fig.~#1\null}
\def\tableautorefname~#1\null{Table~#1\null}

\usepackage{graphicx}
\usepackage{subfigure}
\usepackage{amsmath}
\usepackage{amssymb}
\usepackage{xcolor}
\usepackage{braket}
\definecolor{beppe}{RGB}{255,0,0}
\definecolor{francesco}{RGB}{0,255,0}
\definecolor{massimiliano}{RGB}{0,0,255}

\usepackage{bm}

\def\r{\textbf{r}}
\def\rp{\dot{\textbf{r}}}

\newcommand{\avg}[1]{\left\langle#1\right\rangle}

\begin{document}
\vspace*{1.5cm}
{\noindent\Huge\sffamily\bfseries  Work Fluctuations in the Active Ornstein-Uhlenbeck Particle model} \vspace*{1cm}

\begin{flushright}
\begin{minipage}[b]{0.85\textwidth}
       {\noindent\Large\sffamily\bfseries Massimiliano Semeraro$^{1*}$, Antonio Suma$^1$, Isabella Petrelli$^1$, Francesco Cagnetta$^2$ and Giuseppe Gonnella$^1$}\vspace{3pt}
       
{$^1$Dipartimento di Fisica, Universit\`a degli Studi di Bari, and INFN, Sezione di Bari, via Amendola 173, 70126 Bari, Italy}\\
{$^2$Institute of Physics, \'Ecole Polytechnique F\'ed\'erale de Lausanne (EPFL) CH-1015 Lausanne, Switzerland}
  
       E-mail: \href{mailto:massimiliano.semeraro@uniba.it}{massimiliano.semeraro@uniba.it}\vspace{10pt}
       
{\noindent\large\sffamily\bfseries Abstract.} 
We study the large deviations of the power injected by the active force for an Active Ornstein-Uhlenbeck Particle (AOUP), free or in a confining potential. For the free-particle case, we compute the rate function analytically in $d$-dimensions from a saddle-point expansion, and numerically in two dimensions by {\it a)} direct sampling of the active work in numerical solutions of the AOUP equations and {\it b)} Legendre-Fenchel transform of the scaled cumulant generating function obtained via a cloning algorithm. 
The rate function presents asymptotically linear branches on both sides and it is independent of the system's dimensionality, apart from a multiplicative factor. For the confining potential case, we focus on two-dimensional systems and obtain the rate function numerically using both  methods  {\it a)} and {\it b)}. We find a different scenario for harmonic and anharmonic potentials: in the former case, the phenomenology of fluctuations is analogous to that of a free particle, but the rate function might be non-analytic; in the latter case the rate functions are analytic, but fluctuations are realised by entirely different means, which rely strongly on the particle-potential interaction. Finally, we check the validity of a fluctuation relation for the active work distribution. In the free-particle case, the relation is satisfied with a slope proportional to the bath temperature. The same slope is found for the harmonic potential, regardless of activity, and for an anharmonic potential with low activity. In the anharmonic case with high activity, instead, we find a different slope which is equal to an effective temperature obtained from the fluctuation-dissipation theorem.

\end{minipage}
\end{flushright}
%\hfill

%\vfill	% to go at the bottom of the page
%\rule{1.0\texwidth}
\noindent\hrulefill
\tableofcontents                                % Makes Table of Contents
\noindent\hrulefill
%\rule{1.0\textwidth}

\newpage	% set now the document features (rules, headers etc.)
\footruleheight{0.5pt}
%\headruleheight{1pt}
%\lhead{}
\rhead{\small\sffamily\bfseries Scaling limits of the active interface equations}
%\lfoot{}
\cfoot{}
\rfoot{\thepage}

\section{Introduction}\label{sec:intro}

An {\it Active Particle} is a physical entity able to transform energy from the environment or an internal reservoir into directed motion \cite{marc1, marc2, rams1, rams2}. These particles are the fundamental constituents of {\it active matter}, a special class of out-of-equilibrium systems which have taken centre stage of statistical mechanics in the last few years. Nature offers already a plethora of examples of active matter systems, as colonies of microorganisms \cite{marc1, bact, bact2}, living cells \cite{marc1, marc2, cells1, cells2}, swarms, schools and flocks~\cite{locust, marc1, marc2, birds}, but active particles can also be produced artificially~\cite{rams1, rams3, rods1, lightact, lightact2}. From a theoretical standpoint, the interest towards active systems was fueled by the display of intriguing collective properties, both on average, such as Motility-Induced Phase Separation~\cite{rtpex, ginelli1, mips1, mips2, catesMIPS} and at the level of fluctuations~\cite{gonn1, nemotoLARGEDEV, gonn4, ketaLARGEDEV, mori}. In the present paper, we focus on the problem of a single active particle free or interacting with an external potential. We take the perspective of thermodynamics of trajectories~\cite{lecomte2007thermodynamic} and study the statistics of path-dependent observables: studies of this kind might reveal interesting properties even for the simplest example of close-to-equilibrium dynamics~\cite{farago}, while offering the possibility for a systematic treatment of arbitrarily far-from-equilibrium systems such as active systems~\cite{dabelowAOUP}.

The dynamics of an (overdamped) active particle under the action of an arbitrary potential $U(\bm{r}(t),t)$ and immersed in a thermal bath with friction coefficient $\gamma$ and temperature $T$ is prescribed by a stochastic equation of motion of the following form:
\begin{equation}\label{eq:AOUP1}
 \dot{\bm{r}}(t) = \bm{v}(t) -\gamma^{-1}\bm{\nabla} U(\bm{r}(t),t) + \sqrt{2 D_T}~\bm{\xi}(t),
\end{equation}
where $\bm{r}(t)$ is the position of the particle and $\bm{v}(t)$ its self-propulsion velocity. The constant $D_T=\gamma^{-1}k_B T$ denotes the translational diffusion coefficient, while $\bm{\xi}(t)$ is a zero-mean and delta-correlated white noise. The statistics of $\bm{v}(t)$ depends on the specifics of the self-propulsion mechanism and several possibilities have been considered in the literature: Run-and-Tumble particles, where the modulus of $\bm{v}(t)$ is constant while the direction changes at random Poissonian times, or Active Brownian Particles, where the direction changes as a Brownian motion, to name but a few examples~\cite{rtp1, rtp2, rtp3, rtp4}. In this paper we adopt the Active Ornstein-Uhlenbeck Particle (AOUP) model, where $\bm{v}(t)$ itself is an Ornstein-Uhlenbeck process~\cite{szamelAOUP, martinAOUP}. The AOUP model retains the fundamental property of active particles, i.e. persistence of motion due to self-propulsion, and has enjoyed recent analytical insights regarding collective properties~\cite{fodorHOWFAR, martinAOUP} and energetics~\cite{capriniAOUP, dabelowAOUP}. Besides, the AOUP model can also represent passive tracers immersed in a bath of active particles, such as a bacterial bath~\cite{koumakis2014directed, maggiactivebaths}. 

The observable we choose to characterise the dynamics of the system of interest is the energy injected by the self-propulsion force, or {\it active work},
\begin{equation}\label{eq:active-work}
W_{\tau} = \int_0^\tau dt\, \gamma\bm{v}(t)\cdot \dot{\bm{r}}(t).
\end{equation}
For a single active particle, the active work can be identified with the heat flowing into the thermal bath~\cite{sekimoto1997kinetic} and is proportional to the entropy production~\cite{gonn1, shankarENTROPY, grandpreENTROPY}, provided the self-propulsion velocity is assumed to be even under time reversal \cite{dabelowAOUP, capriniAOUP, ketaLARGEDEV}.
The active work is a natural observable of interest for the stochastic thermodynamics of active systems~\cite{mandalENTROPY, pietzonkaENTROPY, dabelowAOUP, ekehTHERMO}, since it measures how efficiently active driving is converted into motion. Moreover, the fluctuations of the active work have been shown to be deeply connected to structural and dynamical properties of active systems~\cite{gonn1, whitelamLARGEDEV, gonn4} and studying their large deviations provides a pathway to control the collective behaviour of systems of active particles~\cite{nemotoLARGEDEV, tociuLARGEDEV, cagnettaLARGEDEV, fodorLARGEDEV,ketaLARGEDEV}. 
%Our approach follows the method adopted in~\cite{farago} for the analysis of the power injected by thermal noise on a free Brownian particle, extending the method to the correlated noise which drives AOUPs. In particluar,

We study the asymptotic fluctuations of $W_{\tau}$ via the rate function of the active work,
\begin{equation}\label{eq:rate-function-def}
 I(w) = \lim_{\tau\to \infty} \frac{1}{\tau} \log\left({\textrm{Prob.}\left\lbrace W_\tau/\tau = w \right\rbrace}\right),
\end{equation}
for an AOUP which is free or confined by a potential. On the one hand, our work is motivated by some recent results \cite{gonn4, gonn1,nemotoLARGEDEV,ketaLARGEDEV} showing that, for a variety of active particle models, interaction-related effects such as the formation of clusters~\cite{mips4, mips5, mips6} and the consequent drag against the direction of the active force~\cite{gonn4, gonn1} induce singularities in the active work distribution.
%Therefore, our interest towards confining potentials is stimulated by the possible appearance of similar singular behaviours in the simplest case of a single confined active particle.
On the other hand, we are interested in characterising the free AOUP case and compare it to the corresponding passive problem, where the rate function of the energy injected by the thermal noise displays singular behaviour~\cite{farago}.
The main results of our analysis are summarised below.
\begin{itemize}
 \item We compute $I(w)$ exactly for a free AOUP in arbitrary dimension and find it to display two asymptotically linear tails but no non-analyticities, at variance with the power injected by uncorrelated thermal noise in absence of activity~\cite{farago}. This result is based on the evaluation of sub-exponential contributions which are responsible for the singularity in the case of~\cite{farago}. 
% \item We provide two independent numerical estimates of $I(w)$ for a harmonically confined AOUP which suggest the possible presence of linear tails similarly to the free AOUP;
\item  We estimate numerically $I(w)$ for an AOUP moving in a harmonic and two different anharmonic potential wells, and discuss the phenomenology of rare trajectories coming from lower- or higher-than-average fluctuations. In the harmonic case, we find, within numerical accuracy, the presence of linear tails, and a similar phenomenology to that of the free AOUP. In the anharmonic potentials, we do not find singular behaviour of the rate function in spite of accumulation effects at the well boundaries~\cite{aoup1};% and of the presence of a transition found in literature for the rate function of the particle position \cite{aoup1};
 \item We use analytical results for the free-particle case and numerical results both for the free and the confined problem to discuss the validity of a fluctuation relation \cite{seifert_fl_th} for the active work. We find it to hold for free (exactly) and harmonically confined (numerically) AOUP with respect to the bath temperature, independently of the strength of the active force. For an anharmonically confined particle at high activity, we find instead that a fluctuation relation is satisfied with respect to a different temperature, whose value agrees with the effective temperature obtained from the fluctuation-dissipation relation \cite{let_temp, isa_teff1, isa_teff2}.
 %, a property which connects the probability of realisation of positive and negative fluctuations of a stochastic thermodynamic observable to the temperature $T$ of the thermal bath
\end{itemize}

The remainder of the paper is organised as follows. In~\autoref{sec:free}, we consider a free AOUP and show the calculation of the active work rate function.  
Our derivation, shown in \autoref{ssec:scgf} and \autoref{ssec:saddle-point}, is based on a path-integral calculation of the scaled cumulant generating function (SCGF), which yelds the rate function under Legendre-Fenchel transform. Such an approach yields also preasymptotic corrections to the cumulant generating function, which are generally required in order to guarantee that the rate function coincides with the Legendre-Fenchel transform of the SCGF.
%\maxx{In~\autoref{sec:free}, we consider a free AOUP and show the calculation of the active work rate function.} Our derivation based on path integral techniques  \maxx{shown  in \autoref{ssec:scgf} and \autoref{ssec:saddle-point}, yields also preasymptotic corrections (generally required to be known in order to guarantee that the rate function coincides with the Legendre-Fenchel transform of the scaled cumulant generating function) not present in the calculation for the SCGF of $W_\tau$ given in in~\cite{grandpreENTROPY}, obtained  by solving the eigenvalue problem associated with the free AOUP dynamics.}
In \autoref{ssec:cloning} we provide two independent numerical estimates of the rate function that can be directly compared to our theoretical calculations: one obtained by a direct measure of $I(w)$
% the infinite-$\tau$ limit of $\log\left(\textrm{Prob.}\left\lbrace W_\tau/\tau = w \right\rbrace\right)$ 
from numerical integration of the equations of motion, the other obtained by estimating first the generating function of $W_\tau$'s cumulants with a cloning algorithm~\cite{cloning1, cloning2, cloning3} and then performing a  Legendre-Fenchel transformation (see \autoref{app1:cl_alg} for details).
%
%Let us notice here that the scaled cumulant generating function of $W_\tau$ for a free AOUP in $2$ dimensions has been computed recently in~\cite{grandpreENTROPY}, by solving the eigenvalue problem associated with the free AOUP dynamics. 
%In this respect, our work provides an alternative derivation based on path integral techniques  \maxx{that, as explained in detail in \autoref{ssec:scgf} and \autoref{ssec:saddle-point}, yields also preasymptotic corrections. The expression of such contributions are in general required to be known in order to guarantee that the rate function coincides with the Legendre-Fenchel transform of the scaled cumulant generating function.} In \autoref{ssec:cloning} we provide two independent estimates of the rate function that can be directly compared to our theoretical calculations: one obtained by a direct measure of the infinite-$\tau$ limit of $\log\left(\textrm{Prob.}\left\lbrace W_\tau/\tau = w \right\rbrace\right)$ from numerical integration of the equations of motion, the other obtained by estimating first the generating function of $W_\tau$'s cumulants with a cloning algorithm~\cite{cloning1, cloning2, cloning3} and then performing a  Legendre-Fenchel transformation \maxx{(see \autoref{app1:cl_alg} for details). 
The problem of a confined AOUP is discussed in~\autoref{sec:potential}.
%Here we rely only on the two aforementioned numerical techniques to estimate the rate function. 
Three different confining potentials are considered: a harmonic potential (\autoref{ssec:harmonic}), a `stiff' potential growing as $r^{10}$, with $r$ the distance of the particle from the origin, and a circular-shaped rigid barrier of fixed radius whose borders are modelled trough a WCA potential (\autoref{ssec:nonlinear}). 
%For each of these potentials we describe the properties of trajectories conditioned on rare fluctuations of the active work, highlighting how the non-equilibirum accumulation at the boundary of anharmonic potentials \cite{aoup1} induces different means of producing large fluctuations with respect to the free and harmonic cases.
\autoref{sec:fl_th} is devoted to the study of fluctuation relations for the active work in three settings: the free-particle case, the harmonically confined AOUP and the AOUP confined by the `stiff' anharmonic potential. Finally, in \autoref{sec:conclusions}, we report the conclusions and final comments of our study.

 %\maxx{we report a  In particular, using the analytical results from \autoref{sec:free}, we show that the free-particle distribution satisfies a detailed fluctuation theorem with respect to the bath temperature $T$. Two proofs are provided, one based on the check of the symmetry property of the scaled cumulant generating function, and another based on direct calculation involving the rate function. Both proofs are supported by numerical results. We also numerically show that the distribution satisfies the same fluctuation theorem in the harmonically confined case, both with low and high activity, and in the anharmonically confined case at low activity. The high-activity anharmonically confined case instead satisfies a fluctuation theorem, but, differently from the previous cases, with respect to an effective temperature. This effective temperature is evaluated directly from the fluctuation theorem data and found in agreement with an independent estimate obtained in terms of the violation of the equilibrium fluctuation-dissipation relation involving the mean square displacement of the particle and the integrated linear response function (see \autoref{app2:teff_est} for details). }

\section{Active Work of the free Active Ornstein-Uhlenbeck particle}
\label{sec:free}

Let us begin by recalling the details of the overdamped AOUP model. The model consists of two equations, for the $d$-dimensional position $\bm{r}(t)$ and self-propulsion velocity $\bm{v}(t)$ of the AOUP. Regarding the parameters appearing in the equations, we follow the convention of~\cite{aoup1} and write
\begin{subequations}\label{eq:AOUP}
 \begin{align}
  \label{eq:AOUP-position} \dot{\bm{r}}(t) &= \bm{v}(t) -\gamma^{-1}\bm{\nabla} U(\bm{r}(t),t) + \sqrt{2 D_T}~\bm{\xi}(t),\\
  \label{eq:AOUP-propulsion} \dot{\bm{v}}(t) &= -\gamma_R \bm{v}(t) + \sqrt{2 D'_R}~\bm{\eta}(t),
 \end{align}
\end{subequations}
 where both $\bm{\eta}(t)$ and $\bm{\xi}(t)$ are zero-mean, unit-variance independent white noises, i.e.
\begin{equation}
\avg{\xi_i(t) \xi_j(t')} = \avg{\eta_i(t) \eta_j(t')} =
\delta_{i,j}\delta(t-t')\,\quad  \forall~i,j=1,\dots,d.\end{equation}
The damping coefficient $\gamma_R$ controls the exponential decay of correlations in $\bm{v}(t)$. Intuitively, $\gamma_R$ can be written as $(d-1)D_R$, where $d$ is the spatial dimension and $D_R$ the rotational diffusion coefficient of the direction of self-propulsion. As a result, the correlations of the self-propulsion satisfy
\begin{equation}\label{eq:self-propulsion-correlation}
 \avg{\bm{v}(t)\cdot \bm{v}(t)} \xrightarrow{|t-t'|\to\infty} \frac{d D'_R}{(d-1)D_R} e^{-(d-1)D_R|t-t'|}.
\end{equation}
The coefficient $d D'_R / (d-1) D_R$ appearing in the right-hand side of~\autoref{eq:self-propulsion-correlation} is the square of the typical modulus of the self-propulsion velocity, which can be thought of as the ratio between a typical self-propulsion force $F_a$ and the mobility $\gamma$. In other words, the parameter $D'_R$ is fixed by $ d D'_R\,{=}\, (d-1) D_R (F_a/\gamma)^2$. The translational diffusion coefficient $D_T$ is given by $D_T\,{=}\, \gamma^{-1}k_BT$ and the rotational diffusion coefficient is also proportional to the thermodynamic temperature $k_BT$. For a disk-shaped particle in $d\,{=}\,2$, for instance, $D_R\,{=}\, 3 D_T/\sigma^2$, with $\sigma$ the particle diameter. In general, $D_T$ and $D_R$ could be considered as independent parameters, incorporating both thermal and active fluctuations: our choice implies a purely thermal origin, so that the only source of departure from equilibrium is the self-propulsion velocity. To sum up, the free parameters of the model are the temperature $T$, the typical magnitude of the active force $F_a$ and the particle mobility $\gamma$. Additional parameters, required in order to specify the potential, will be introduced in~\autoref{sec:potential}. In the remainder of this section, we set the potential $U(\bm{r}(t),t)$ to $0$ and compute the large deviations of the active work (\autoref{eq:active-work}) for the free AOUP.

Let us then turn to the main focus of this manuscript, the asymptotics of the probability-density-function (pdf) $\Pi(w)$ of the active work, 
\begin{equation}\label{eq:active-work-pdf}
 \Pi(w) = \avg{\delta(W_\tau-\tau w)} \asymp e^{-\tau I(w)},
\end{equation}
where $W_\tau$ denotes the active work as a random variable and $w$ the specific realisations, scaled by the obervation time $\tau$. The average here is performed over realisations of the stochastic noises $\bm{\xi}(t)$ and $\bm{\bm{\eta}}(t)$; the symbol $\asymp$ denotes equality of the large-$\tau$ limit on the logarithmic scale~\cite{touch, ellis3}. By introducing, as it is customary, the Laplace representation of the delta function, $\delta(x) = (1/2\pi i) \int_{-i\infty}^{i\infty} d\lambda\, e^{\lambda x}$, we can write
\begin{equation}\label{eq:anti-laplace}
 \Pi(w) = \frac{1}{2 \pi i}\int_{-i\infty}^{i\infty} d\lambda\, e^{-\tau \lambda w}\avg{ e^{\lambda W_{\tau}}} \equiv \frac{1}{2 \pi i}\int_{-i\infty}^{i\infty} d\lambda\, e^{-\tau \lambda w}\hat{\Pi}(\lambda),
\end{equation}
where we have introduced the generating function of $W_{\tau}$'s cumulants (CGF), $\hat{\Pi}(\lambda)$. Because of the exponential factor in the integrand, once $\hat{\Pi}(\lambda)$ is obtained, the asymptotic of $\Pi(w)$ can be estimated with a saddle-point expansion. We now turn to the computation of the cumulant generating function $\hat{\Pi}(\lambda)$. The saddle-point expansion is performed in \autoref{ssec:saddle-point}. Finally, in~\autoref{ssec:cloning}, we compare our analytical prediction with the result of direct simulations of the model and a refined estimate obtained with a biased sampling of the model trajectories.

\subsection{The cumulant generating function}\label{ssec:scgf}

In this section we compute the generating function of $W_{\tau}$'s cumulants,
\begin{equation}\label{eq:active-work-cgf}
 \hat{\Pi}(\lambda) = \avg{ e^{\lambda W_{\tau}}},
\end{equation}
for the free AOUP, following the approach of ref.~\cite{farago}. The average appearing in \autoref{eq:active-work-cgf} is performed with respect to the realisations, or paths, of the noises $\bm{\xi}(t)$ and $\bm{\eta}(t)$ affecting position and self-propulsion speed in~\autoref{eq:AOUP}. Since, by~\autoref{eq:AOUP-position},
\begin{equation}
 W_{\tau} = \gamma\int_0^\tau dt\, \bm{v}(t)\cdot\rp(t) = \gamma\int_0^\tau dt\, \bm{v}(t)\cdot\bm{v}(t) + \gamma\sqrt{2 D_T} \int_0^\tau dt\, \bm{v}(t)\cdot \bm{\xi}(t),
\end{equation}
the path probabilities required to perform the average are those of $\bm{\xi}(t)$ (denoted $\mathcal{P}[\bm{\xi}(t)]$) and $\bm{v}(t)$ (denoted $\mathcal{P}[\bm{v}(t)|\bm{v}_0] p_0(\bm{v}_0)$, conditioned to the initial condition $\bm{v}_0$ with initial probability $p_0(\bm{v}_0)$). These path probabilities refer to time intervals of length $\tau$, but we omit $\tau$ from the symbols so as to ease the notation. The relevant path probabilities are given by~\cite{paths},
\begin{equation}\label{eq:path-probabilities}
\begin{aligned}
	\mathcal{P}[\boldsymbol{\xi}(t)]& = \exp\left\{-\frac{1}{2}\int_0^\tau dt~\boldsymbol{\xi}(t)\cdot\boldsymbol{\xi}(t)\right\},\\
	\mathcal{P}[\bm{v}(t)|\bm{v}_0]&=\exp\left\{-\frac{1}{4D'_R}\int_0^\tau dt\,(\dot{\bm{v}}(t)+\gamma_R\bm{v}(t))^2+\frac{d}{2}\gamma_R\tau\right\},
\end{aligned}
\end{equation}
where the time-integrals are to be intended according to the Stratonovich discretisation scheme~\cite{gardiner1985handbook}. The corresponding measures, $\mathcal{D}[\boldsymbol{\xi}(t)]$ and $\mathcal{D}[\bm{v}(t)|\bm{v}_0]$ include also the proper normalisation factor. $\mathcal{P}[\bm{v}(t)|\bm{v}_0]$, in particular, is the path probability of the self-propulsion process with fixed endpoints $\bm{v}_0$ and $\bm{v}(\tau)\equiv \bm{v}_\tau$. Denoting with $p(\bm{v}_\tau,\tau|\bm{v}_0,0)$ the transition probability of the Ornstein-Uhlenbeck process~\cite{gardiner1985handbook}, the normalisation factor is thus fixed by the following condition,
\begin{equation}
 \int \mathcal{D}[\bm{v}(t)|\bm{v}_0]\mathcal{P}[\bm{v}(t)|\bm{v}_0] \,{=}\, p(\bm{v}_\tau,\tau|\bm{v}_0,0).
\end{equation}
We can now unwind the average symbol of~\autoref{eq:active-work-cgf}, so as to have
\begin{equation}\label{eq:cgf1}
\begin{aligned}
\hat{\Pi}(\lambda)&=\int d\bm{v}_0\,d\bm{v}_\tau\,\mathcal{D}[\boldsymbol{\xi}(t)]\mathcal{D}[\bm{v}(t)|\bm{v}_0]\mathcal{P}[\bm{v}(t)|\bm{v}_0]\mathcal{P}[\boldsymbol{\xi}(t)]\,p_0(\bm{v}_0)\times \\
&\exp\left\{
\lambda \gamma\left(
\int_0^\tau dt\,\bm{v}(t)\cdot\bm{v}(t)+\sqrt{2D_T}\int_0^\tau dt\,\bm{v}(t)\cdot\boldsymbol{\xi}(t)
\right)\right\},
\end{aligned}
\end{equation}
which satisfies the normalisation condition $\hat{\Pi}(0)\,{=}\,1$.

Because of the Stratonovich convention, the rules of standard calculus can be used to simplify the right-hand side of~\autoref{eq:cgf1}:
\begin{equation}\label{eq:cgf2}\begin{aligned}
 \hat{\Pi}(\lambda) =\, & e^{\frac{d\gamma_R\tau}{2}} \int d\bm{v}_0 p_0(\bm{v}_0)\, d\bm{v}_\tau\,\mathcal{D}[\bm{v}(t)|\bm{v}_0]\, \mathcal{D}[\boldsymbol{\xi}(t)]\exp\left\{-\frac{\gamma_R}{4D'_R}(\bm{v}_\tau^2-\bm{v}_0^2)\right\}\times\\
&\exp\left\{
-\frac{1}{4D'_R}\int_0^\tau dt \left(\dot{\bm{v}}^2(t)+\alpha^2\bm{v}^2(t)\right) -\frac{1}{2}\int_0^\tau\left(\boldsymbol{\xi}(t)-\sqrt{2D_T}\lambda\gamma\bm{v}(t)\right)^2
\right\},
\end{aligned}\end{equation}
where we have set $\alpha^2 \,{=}\,\gamma_R^2 - 4 D'_R \lambda \gamma(1+\lambda \gamma D_T)$. After the linear change of variables $\bm{\xi}'(t)\,{=}\,\bm{\xi}(t)-\sqrt{2D_T}\gamma\lambda\bm{v}(t)$ the integral over the thermal noise can be performed---it equals $1$ due to normalisation. The remaining terms can all be written as products over the spatial components of $\bm{v}(t)$, so that the integral, due to the independence of the components of an Ornstein-Uhlenbeck process, factorises. By isotropy, each of the factors yields the same contribution. We will, in addition, make use of the following identity for a one-dimensional Ornstein-Uhlenbeck process $v(t)$ and $\alpha\in \mathbb{R}$~\cite{pathintegral},
\begin{equation}\label{eq:cgf3}\begin{aligned}
&\int \mathcal{D}[{v}(t)|v_0] \exp\left\{-\frac{1}{4D'_R}\int_0^\tau\left[\dot{v}(t)^2+\alpha^2 v(t)^2\right]\right\} = \\
&= \left( \pi\frac{4D'_R}{\alpha}\sinh(\alpha\tau)\right)^{-\frac{1}{2}}\exp\left\{-\frac{\alpha}{4D'_R}\frac{(v_\tau^2+v_0^2)\cosh(\alpha\tau)-2v_0\,v_\tau}{\sinh(\alpha\tau)}\right\}\,,
\end{aligned}
\end{equation}
and set $p_0(v_0)$ to the stationary probability of the self-propulsion process, 
\begin{equation}\label{eq:init-prob}
 p_0( v_0 ) = \sqrt{\frac{\gamma_R}{2\pi D'_R}} e^{-\frac{\gamma_R}{2 D'_R} v_0^2}.
\end{equation}
Therefore, after performing the integrals over $\bm{v}_0$ and $\bm{v}_\tau$, we obtain the following expression for the cumulant generating function,
\begin{equation}\begin{aligned}\label{eq:cgf4}
\hat{\Pi}(\lambda) &= e^{\frac{d\gamma_R\tau}{2}}\left(\frac{\gamma_R^2+\alpha^2}{\alpha\gamma_R}\sinh(\tau\alpha)+2\cosh(\tau \alpha)\right)^{-\frac{d}{2}} \\
&= \frac{e^{\tau\frac{d}{2}\left(\gamma_R - \alpha\right)}}{\left(\frac{1 + e^{-2\tau\alpha}}{2}\right)^{d/2}\left( 2 + \frac{\gamma_R^2 + \alpha^2}{\gamma_R \alpha}\tanh(\tau\alpha)\right)^{d/2}}.
\end{aligned}\end{equation}
Let us pause briefly and comment on the last result: the cumulant generating function consists of a factor which grows exponentially with $\tau$ and a sub-exponential prefactor. The exponential factor gives rise to the {\it scaled} cumulant generating function (SCGF),
\begin{equation}\label{eq:SCGF_def}
 g(\lambda) = \lim_{\tau\to\infty}\frac{1}{\tau} \log{\hat{\Pi}(\lambda)},
\end{equation}
which reads, for this specific problem,
\begin{equation}\label{eq:SCGF}
 g(\lambda) = \frac{d}{2}\left(\gamma_R -\sqrt{\gamma_R^2 - 4 D'_R \gamma \lambda (1+ \gamma D_T \lambda)} \right).
\end{equation}
This expression agrees with the earlier result of~\cite{grandpreENTROPY} in $d\,{=}\,2$ and generalises it to {\it any} spatial dimension. Importantly, we also notice that if the parameters $\gamma_R$ and $D'_R$ were free to take any value, instead of being constrained by the relations expounded at the beginning of the section, then the spatial dimension would only appears as an overall multiplicative factor in $g(\lambda)$. The implication is that, for the free AOUP, increasing the spatial dimension only increases the {\it rate} at which the probability $\Pi(w)$ of active work fluctuations concentrates around the mean, while it does not affect the overall shape of the rate function.

In general, the leading exponential behaviour of $\hat{\Pi}(\lambda)$ provides enough information for the calculation of the rate function via Legendre-Fenchel transform~\cite{ellis3, touch}. However, a complex subexponential prefactor such as the one appearing on the right-hand side of \autoref{eq:cgf4} might result in additional non-analyticities of $\hat{\Pi}(\lambda)$ which still affect the leading exponential behaviour~\cite{farago}. This property is clear when the inverse Laplace transform in~\autoref{eq:anti-laplace} is computed with a saddle-point method: denoting with $F(\lambda)$ the subexponential prefactor, such that $\hat{\Pi}(\lambda) \,{=}\, F(\lambda) e^{\tau g(\lambda)}$,~\autoref{eq:anti-laplace} can be cast in the following form,
\begin{equation}\label{eq:anti-laplace-saddle}
 \Pi(w) = \frac{1}{2 \pi i}\int_{-i\infty}^{i\infty} d\lambda\, F(\lambda)\, e^{-\tau\left[ \lambda w - g(\lambda)\right]}.
\end{equation}
Not only $F(\lambda)$ concurs to the subexponential corrections to the asymptotics of $\Pi(w)$, but also its non-analyticities might pose severe limitations to the deformation of the integration contour. Therefore, it is worth include the subexponential prefactor in the saddle-point estimation of the integral above, which is carried out in the next section.

\subsection{Saddle-point estimation of the rate function}\label{ssec:saddle-point}

\begin{figure}[t!]
	\centering
	\includegraphics[width=0.7\textwidth]{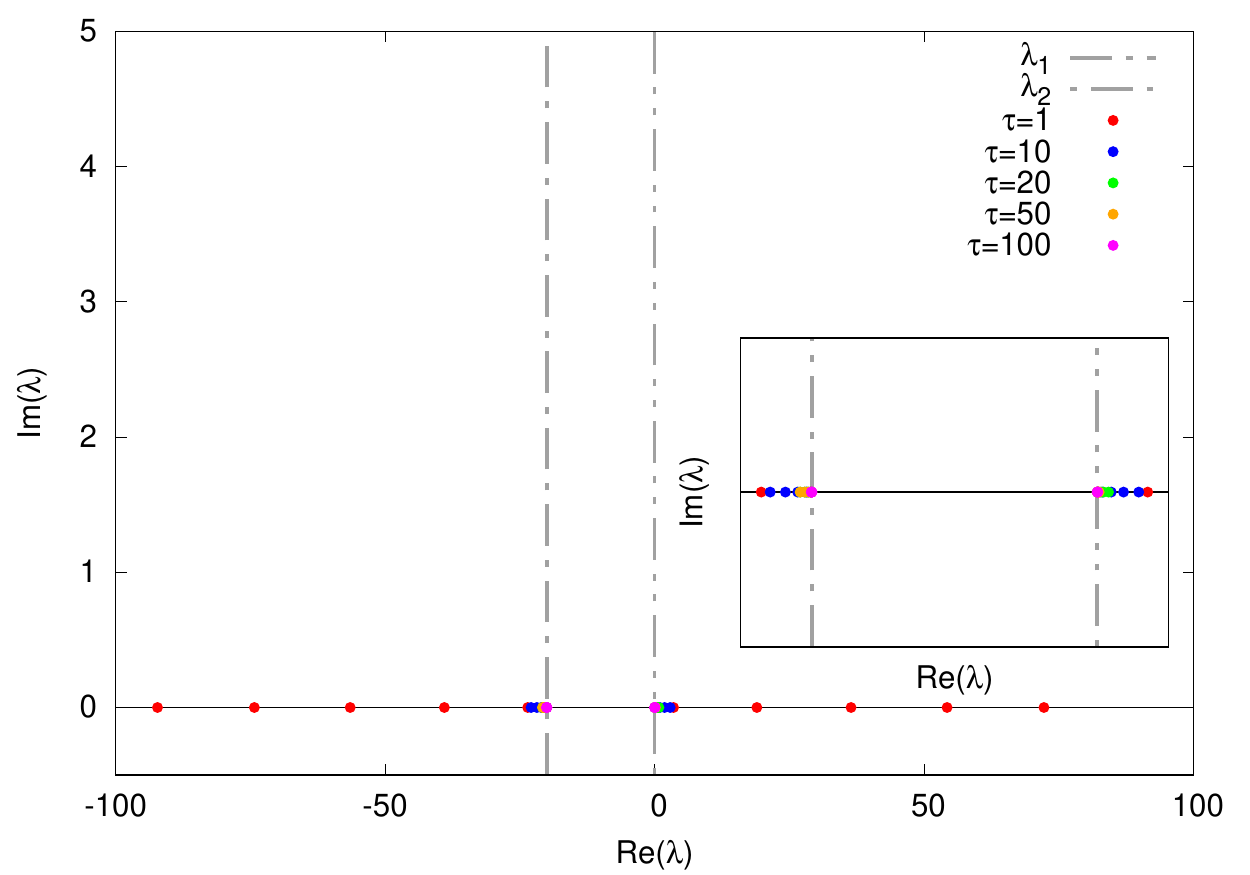}
	\caption{\footnotesize{Poles of $F(\lambda)$ (\autoref{eq:sing-F}) in the complex plane, for values of $\tau$ as in the caption. Here $\gamma\,{=}\,10$, $F_a\,{=}\,20$ and $k_BT\,{=}\,0.05$. The two vertical dot-dashed lines mark the position of the two branch points caused by the structure of $g(\lambda)$, respectively at $\lambda_1\sim-20.0003$ and at $\lambda_2\sim0.0003$. The inset shows an enlargement of the main figure around the interval $[\lambda_1, \lambda_2]$.}}
	\label{fig:non-analiticities}
\end{figure}

In order to simplify the discussion, we introduce the following rescaled variables,
\begin{equation}\begin{aligned}
& \tilde{\gamma}_R = \tau \gamma_R,\quad \tilde{\lambda} = \frac{4 D'_R \gamma}{\gamma_R^2} \lambda, \\ &\tilde\alpha = \frac{\alpha}{\gamma_R} = \sqrt{1 - \tilde{\lambda} - A \tilde{\lambda}^2},\,\text{ with } A =\frac{\gamma_R^2 D_T}{ 4 D'_R}.
\end{aligned}\end{equation}
In terms of the new variables,
\begin{subequations}\label{eq:sing}
\begin{align}
\label{eq:sing-g} g(\lambda) &= \frac{d}{2}\gamma_R (1-\tilde\alpha),\\
\label{eq:sing-F} F(\lambda) &= \left(1+e^{-2\tilde\alpha\tilde\gamma_R}\right)^{-d/2}\left(1+\frac{1}{2}\left(\tilde\alpha+\frac{1}{\tilde\alpha}\right)\tanh(\tilde\gamma_R\tilde\alpha)\right)^{-d/2}.
\end{align}
\end{subequations}
Although the original variables were different, in terms of the rescaled variables we get exactly the same result as~\cite{farago}, hence the same considerations apply here as well. In particular, the square root in the definition of $\alpha$ introduces two branch points at
\begin{equation}\label{eq:cgf-boundary}
\tilde\lambda = \tilde\lambda_{1/2} = -\frac{1 \pm \sqrt{1+4A}}{2A}.
\end{equation}
Assuming $\tilde\alpha\,{>}\,0$ for positive real arguments of the square root (i.e. $\tilde\lambda_{1} \,{<}\,\tilde\lambda \,{<}\, \tilde\lambda_{2}$) is equivalent to considering two branch cuts on the real axis: one originating in $\tilde\lambda_{1}$ and running towards negative values, the other originating in $\tilde\lambda_{2}$ and running towards positive values. Other possible singularities are the poles of $F(\lambda)$ (\autoref{eq:sing-F}). It is straightforward to check that the denominator of $F(\lambda)$ has purely imaginary roots $\tilde\alpha = i y$, with $y$ satisfying
\begin{equation}
 \left(y - \frac{1}{y}\right)\tan\left(\tilde\gamma_R y\right) = 2.
\end{equation}
In terms of $\tilde\lambda$, these points are located at
\begin{equation}
\tilde\lambda_{1/2}(y) = -\frac{1 \pm \sqrt{1+4A\left(1 + y^2\right)}}{2A},
\end{equation}
thus are covered by the two branch cuts (see~\autoref{fig:non-analiticities}). One concludes that, at variance with the passive case examined in~\cite{farago}, the sub-exponential pre-factor of $\hat\Pi(\lambda)$ does not induce additional singularities in the complex plane. Let us notice that the overall picture can be extended to all dimensions $d$, although for odd dimensions the roots of the denominator of~\autoref{eq:sing-F} generate branch points rather than poles. 

Knowing the analyticity of $\hat{\Pi}(\lambda)$ we can now proceed with the calculation of $\Pi(w)$, whose integral expression we repeat here for clarity,
\begin{equation}
 \Pi(w) = \frac{1}{2 \pi i}\int_{-i\infty}^{i\infty} d\lambda\, F(\lambda)\, e^{-\tau\left[ \lambda w - g(\lambda)\right]}.
\end{equation}
The procedure is the following: first, one must identify the stationary points of the function at the exponent, which are all saddle points in the complex plane. Then, by Cauchy's theorem, the integration contour can be deformed so that it passes through the highest of the saddle points. In particular, the deformed contour is chosen so that it crosses the saddle point along the direction of {\it steepest descent}, so that only the portion of the deformed contour close to the saddle point contributes to the integral as $\tau\to\infty$~\cite{orzsag}.

In the present problem, the exponent of the integrand is, from~\autoref{eq:anti-laplace-saddle}, $ f(\lambda)\,{=}\,\lambda w -g(\lambda)$. It is convenient to consider again the rescaled variables, i.e.
\begin{equation}
 f(\tilde\lambda) = \gamma_R \left( \tilde w \tilde\lambda -1 +\tilde\alpha(\tilde\lambda) \right),
\end{equation} 
with $\tilde w = \gamma_R w / (4 D'_R \gamma)$. The condition $f'(\tilde\lambda)\,{=}\,0$ is satisfied by
\begin{equation}
 \tilde\lambda^{(s)}_\pm = -\frac{1\pm \sqrt{1 + 4 A \frac{4 \tilde w ^2 -1}{4(A + \tilde w )^2}}}{2A}
\end{equation}
and it is straightforward to check that the highest saddle point is $\tilde\lambda^{(s)}_-$ for $\tilde w \,{\geq}\, 0$, $\tilde\lambda^{(s)}_+$ for $\tilde w \,{<}\, 0$. It is also worth noticing that the saddle point lies within the interval $\left[\tilde\lambda_1, \tilde\lambda_2\right]$ for every value of $\tilde w$, so that the integration contour can always be deformed so as to pass by the saddle point without encountering any non-analyticity of the integrand.

\begin{figure}[ht!]
\begin{center}
  \begin{tabular}{cc}
       \includegraphics[width=0.5\columnwidth]{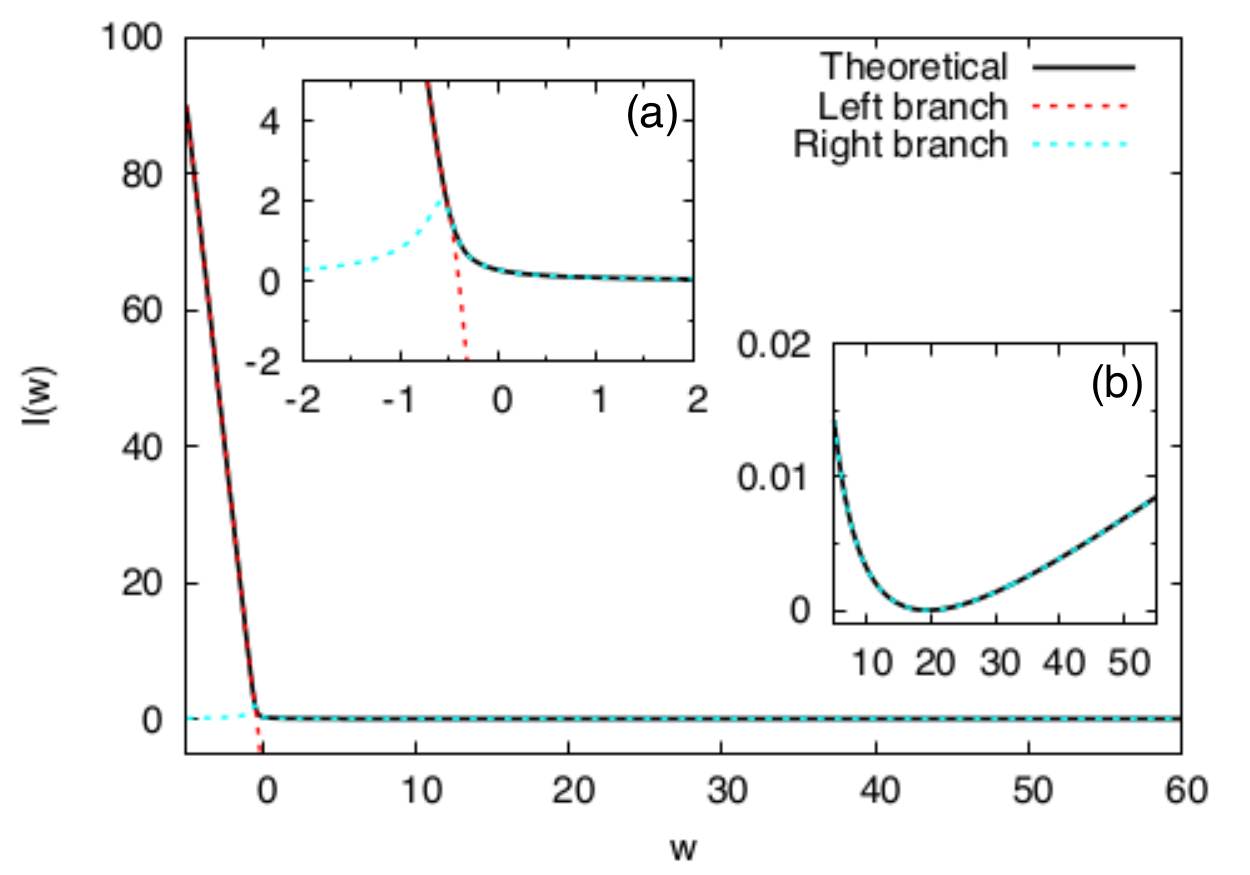}
       \includegraphics[width=0.5\columnwidth]{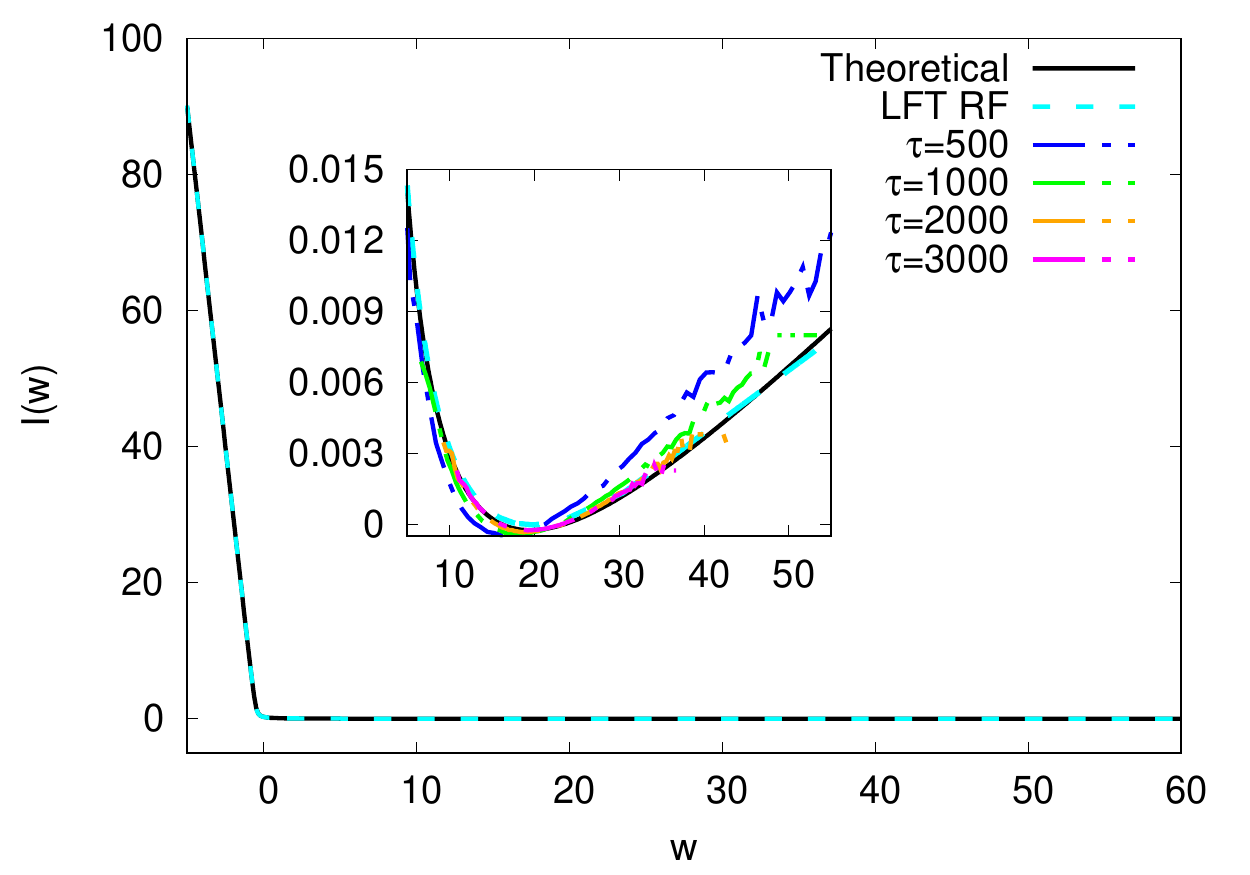}\\
  \end{tabular}
\caption{\footnotesize{Left: Rate function of the free AOUP \autoref{eq:rate-function} in $d\,{=}\,2$, for $\gamma\,{=}\,10$, $F_a\,{=}\,20$ and $k_BT\,{=}\,0.05$ (black solid line). The two dashed lines correspond to the left branch with $\lambda^{(s)}\,{=}\,\lambda^{(s)}_+$ (red) and the right branch with $\lambda^{(s)}\,{=}\,\lambda^{(s)}_-$ (cyan). Inset (a) shows an enlargement of the main figure near $w\,{=}\,0$, where the two branches merge, whereas inset (b) shows an enlargement near the minimum of $I(w)$, namely the average active work $\avg{W_\tau}/\tau$. Right: numerical estimates of the rate function of the active work for the free AOUP obtained via direct numerical integration of~\autoref{eq:AOUP1-overd} using the same parameters of the left panel.  The  theoretical estimate (black solid line) and the Legendre-Fenchel transform of the cloning SCGF from \autoref{fig:free-SCGF-rate-cloning} (cyan dashed line, labelled `LFT SCGF' in the key) are also shown for comparison. The inset shows the same plot restricted to the range of fluctuations which can be reached by direct numerical sampling, generally smaller than the range accessed via the cloning estimate.}}
\label{fig:free-rate-function}
\end{center}
\end{figure}

Although analytical expressions for the steepest descents curves are not available, it can be shown that they cross the saddle point in the direction of the imaginary axis. Therefore, after performing a quadratic expansion of the exponent $f(\tilde\lambda)$ around the saddle point, the following asymptotic expression is obtained for $\Pi(w)$,
\begin{equation}\label{eq:saddle-point-exp}
 \Pi(w) \asymp \frac{F(\tilde\lambda^{(s)})}{2\pi} \left(\frac{2\pi}{ \tau \mathcal{C}}\right)^{1/2} e^{-\tau\gamma_R\left[\tilde w \tilde\lambda^{(s)} -1 + \tilde\alpha(\tilde\lambda^{(s)}) \right] }.   
\end{equation}
Here $\mathcal{C}$ denotes the modulus of the second derivative of $f(\lambda)$ along the steepest descent path---i.e. the imaginary direction in the complex plane---and $\tilde\lambda^{(s)}(\tilde w)$ coincides with $\tilde\lambda^{(s)}_+$ for negative $\tilde w$'s and with $\tilde\lambda^{(s)}_-$ for positive $\tilde w$'s. From the exponential factor on the right-hand side of~\autoref{eq:saddle-point-exp} we can finally extract a definitio, for the rate function of the active work's fluctuations,\begin{equation}\label{eq:rate-function}
I(w) = \frac{d\gamma_R}{2} \tilde I\left(\frac{2}{d} \frac{\gamma_R}{ 4D'_R \gamma} w\right), \quad \tilde I(\tilde w) = \sqrt{\frac{1 + 4A}{A + \tilde w ^2 }}\left(\frac{A + \tilde w^2}{2A}\right) -1 -\frac{\tilde w}{2A},\quad A=\frac{\gamma^2_R D_T}{4D'_R},
\end{equation}
%piecewise definition for the rate function of the active work's fluctuations,\begin{equation}\label{eq:rate-function} I(w) = \left\lbrace \begin{aligned}  &-\gamma_R + w \lambda^{(s)}_-(w) + \sqrt{\gamma_R^2 - 4 D'_R \gamma \lambda^{(s)}_-(w)\left(1 + \gamma D_T \lambda^{(s)}_-(w) \right)},\text{ for } w\geq 0,\\ &-\gamma_R + w \lambda^{(s)}_+(w) + \sqrt{\gamma_R^2 - 4 D'_R \gamma \lambda^{(s)}_+(w)\left(1 + \gamma D_T \lambda^{(s)}_+(w) \right)} ,\text{ for } w< 0. \end{aligned}\right.\end{equation}
%(with $\lambda_{\pm}^{(s)}(w) \,{=}\,\gamma^2 _R \tilde\lambda_{\pm}^{(s)}(w) / 4 D'_R \gamma$),
%\begin{equation}\label{eq:rate-function}
%I(w) = \left\lbrace \begin{aligned}  &-\gamma_R + w \lambda^{(s)}_-(w) + \sqrt{\gamma_R^2 - 4 D'_R \gamma \lambda^{(s)}_-(w)\left(1 + \gamma D_T \lambda^{(s)}_-(w) \right)},\text{ for } w\geq 0,\\ &-\gamma_R + w \lambda^{(s)}_+(w) + \sqrt{\gamma_R^2 - 4 D'_R \gamma \lambda^{(s)}_+(w)\left(1 + \gamma D_T \lambda^{(s)}_+(w) \right)} ,\text{ for } w< 0. \end{aligned}\right.
%\end{equation}
which is shown in~\autoref{fig:free-rate-function}, left panel, for a specific choice of the system parameters. At variance with the rate function of the power injected by the thermal noise on a Brownian particle, which becomes linear abruptly at a certain critical value of $w$~\cite{farago}, the rate function of~\autoref{eq:rate-function-def} is only asymptotically linear. Regarding the strong asymmetry of fluctuations about the mean, however, the phenomenology is similar to the passive case: positive fluctuations of the Ornstein-Uhlenbeck noise generate trajectories with a large velocity, resulting in an even higher energy uptake at later times. Due to this positive feedback, higher-than-average fluctuations of the active work are significantly more likely than lower-than-average ones, consistently with the slope of the right branch of~\autoref{eq:rate-function} being smaller in modulus than the slope of the left branch.

\subsection{Numerical estimates via direct sampling and cloning algorithm}\label{ssec:cloning}

The rate function of $W_{\tau}$---or any dynamical observable---can also be estimated numerically in two different ways. The first one involves the direct sampling of a large number of values for the active work, obtained by solving the dynamics of the problem for several initial conditions and realisation of the stochastic forces, then applying the definition~\autoref{eq:rate-function-def} to the empirical distribution function. An empirical estimate is in fact all that can be obtained for the general confined AOUP problem, considered in the next section. The second method we employ is the {\it population Monte Carlo}, or simply {\it cloning} algorithm~\cite{cloning1, cloning2, cloning3}, where one samples the scaled cumulant generating function and then obtains the rate function using the Legendre-Fenchel transform. This method is especially useful to access values of the active work in the tails of the distribution, which typically require a number of samples exponentially large in $\tau$. The details of both numerical integration methods are reported in \autoref{app1:cl_alg}. Here we only report our choice of parameters: we set $k_BT\,{=}\,0.05$, $\gamma\,{=}\,10$ and $\sigma\,{=}\,1$, such that $D_T\,{=}\,0.005$ and $D_R=0.015$. Moreover we set the self-propulsion force to $F_a\,{=}\,20$, such that the P\'eclet number, computed as $Pe\,{=}\, \sigma F_a / k_B T\,{=}\,4\times 10^2$~\cite{aoup1}, is large.

\begin{figure}[ht!]
\begin{center}
  \begin{tabular}{cc}
       \includegraphics[width=0.5\columnwidth]{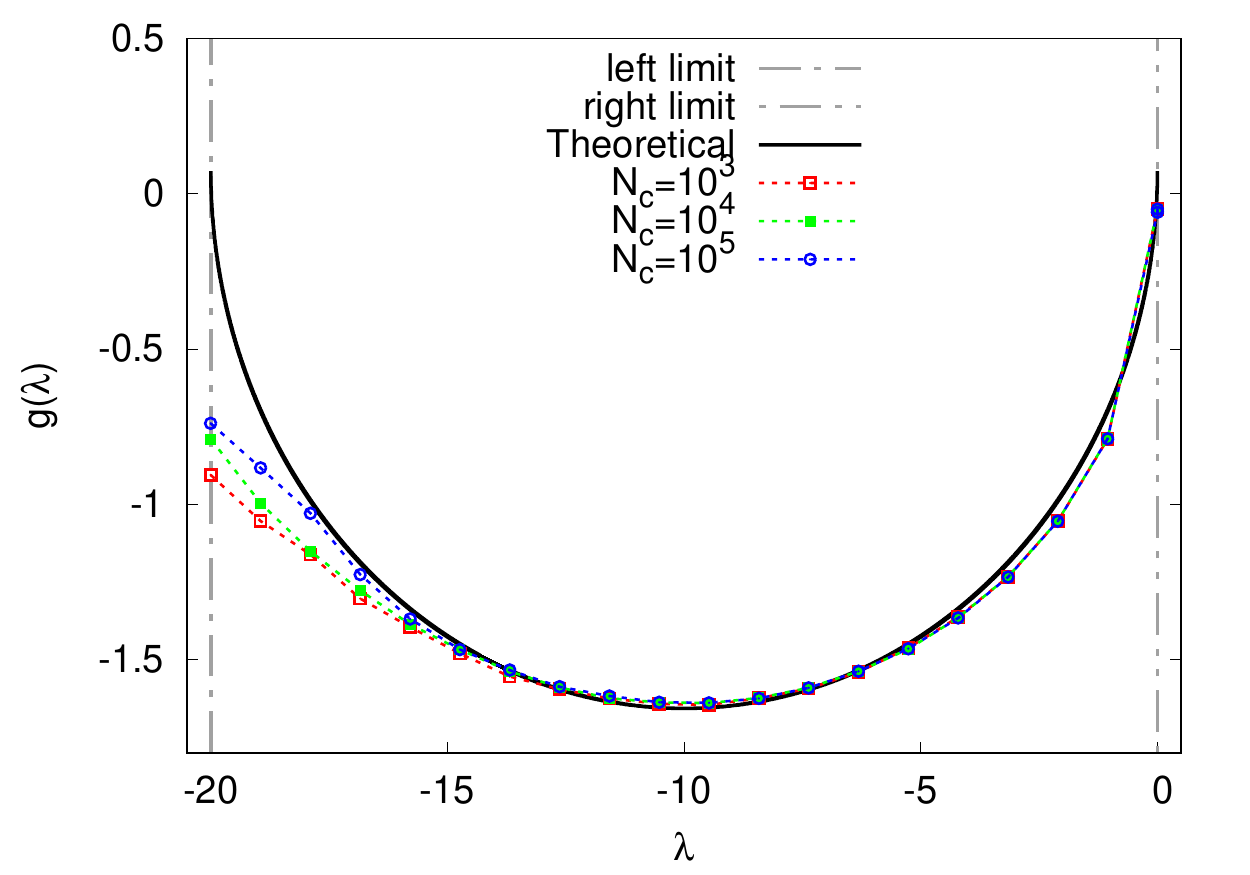}
       \includegraphics[width=0.5\columnwidth]{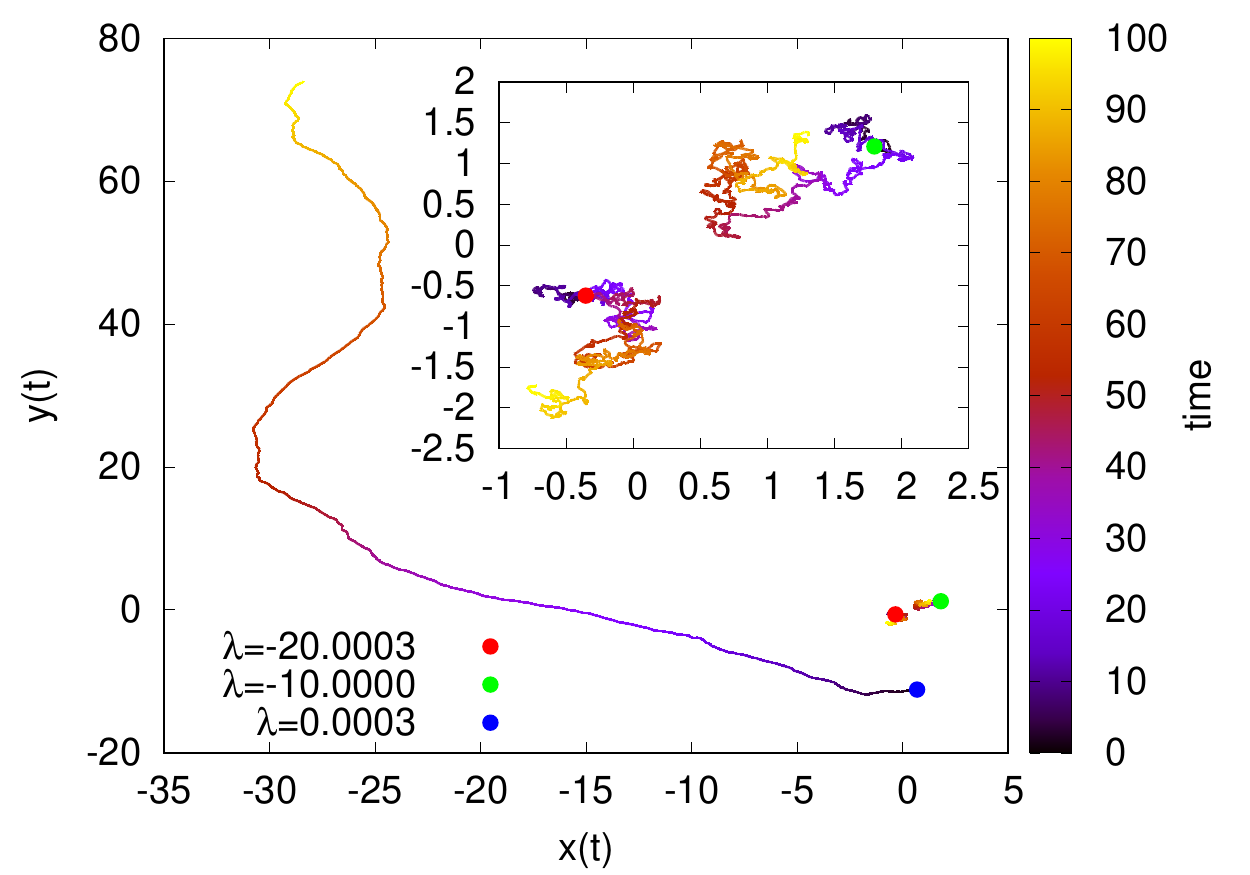}\\
  \end{tabular}
\caption{\footnotesize{Left: SCGF of the free AOUP, numerical (coloured dots) and analytical (black solid line). Parameters are $\gamma\,{=}\,10$, $F_a\,{=}\,20$ and $k_BT\,{=}\,0.05$. For $\lambda$ close to zero the numerical estimate approaches the theoretical curve for a comparatively low number of copies $N_c$ of the cloning algorithm, whereas progressively higher $N_c$'s are required for larger $\lambda$'s. The denser sampling for $N_c=10^3$ is required for the evaluation of the LFT reported in~\autoref{fig:free-rate-function} right panel. Right: Trajectories of the clones for three representative values of $\lambda$: $\sim 0$, close to the right boundary of the domain of $g(\lambda)$; $\sim -20$, close to the left boundary; $\sim -10$, close to the minimum of $g(\lambda)$, which corresponds to vanishing active work on average. Trajectories are coloured to show the evolution of the trajectory in time, according to the colour bar on the right-hand side of the figure. The starting points of the trajectories are sampled from the position distribution of the biased process.}}
\label{fig:free-SCGF-rate-cloning}
\end{center}
\end{figure}

\autoref{fig:free-rate-function} shows the empirical rate functions on the right, obtained from the histogram of the active work using~\autoref{eq:rate-function-def}, for different values of $\tau$ as reported in the key. As soon as the observation time $\tau$ approaches $2\times 10^3$, the rate function converges to the exact result.
\autoref{fig:free-SCGF-rate-cloning}, left panel, shows instead a numerical estimate of the free-particle SCGF for the same set of parameters considered in \autoref{fig:free-rate-function} and increasing number of copies $N_c$. Notice that the closer $\lambda$ to zero, the faster the convergence of the numerical estimate to the theoretical curve. In fact, far away from the `typical' value $\lambda \simeq 0$, which corresponds to an unbiased (or almost unbiased) evolution, the estimate of $g(\lambda)$ depends crucially on one of the clones achieving some rare fluctuation. Besides, it is interesting to notice that choosing a value of $\lambda$ which lies outside of the domain of $g(\lambda)$ results in a break down of the numerical scheme. Specifically, while the numerical estimate of $g(\lambda)$ for $\lambda$ close to the left limit of the domain does not overlap with the analytical curve, it is still sensible. Conversely, when applied out of the domain the algorithm returns either diverging or highly oscillating values. This feature is worth remembering when studying problems without a known analytical solution, such as the cases reported in~\autoref{sec:potential}. The rate function can be finally obtained with a Legendre-Fenchel transform (LFT), provided the SCGF is smooth and {\it steep}~\cite{touch, ellis3}. Note that the accurancy of the LFT depends on the sampling density of the SCGF itself. For instance, \autoref{fig:free-rate-function} right shows a comparison between the LFT of the SCGF and the rate function obtained via direct measurement: in order to achieve high accurancy around the minimum (shown in the inset of the figure) we have performed a denser sampling of the cloning SCGF around $\lambda\,{=}\,0$. The cost of a denser sampling is mitigated by the fact that, close to $\lambda\,{=}\,0$, convergence of the SCGF is achieved with a comparatively low number of clones (cf. ~\autoref{fig:free-SCGF-rate-cloning}). The same sampling criterion will be used in the presence of confining potentials in order to compare the LFT with the rate function measured from direct numerical integration of the dynamics.
% in a provides coherent results only if the corresponding cloning SCGF is densely sampled (see~\autoref{fig:free-SCGF-rate-cloning}) in the region in which the domain and derivative of the rate function and of the scgf are related~\cite{touch}. The inset shows instead direct numerical estimates of the rate function of the active work in the free case obtained with the same choice of parameters as in the left panel compared to the LFT SCGF and to \autoref{eq:rate-function}.} In fact, the curve obtained in this case is in great agreement with both the theoretical one and the one obtained using the first numerical method (\autoref{fig:free-rate-function}, right).

The right panel of~\autoref{fig:free-SCGF-rate-cloning} shows the typical trajectories of the clones for a few different values of the bias $\lambda$. These can be thought of as a representation of the trajectories of a model which is biased so as to have a $\lambda$-dependent average active work $w(\lambda)$, with $w(\lambda)$ given by the saddle-point condition $w(\lambda) \,{=}\,g'(\lambda)$. At $\lambda\simeq 0$ (trajectory starting on the blue dot in the figure), the particle follows the typical AOUP dynamics --- $w(0)$ coincides with the unbiased average active work. For $\lambda$ large and negative, the trajectory (starting on the green and red dots in the figure) produces an active work which is much smaller than the free-AOUP average. Specifically, for the one starting from the green dot, with $\lambda\simeq -10$, the active work is close to zero, whereas the one starting from the red dot produces a negative active work. The trajectories themselves hint at the mechanism by which non-typical active works are produced: a vanishing active work, for instance, can be produced by having a sequence of atypically small kicks from the active Ornstein-Uhlenbeck noise, or by having the thermal delta-correlated noise to act in opposition with the active noise: both cases result in the particle moving much less than on average, as it is the case for the corresponding trajectory displayed in~\autoref{fig:free-SCGF-rate-cloning}. A negative active work, instead, requires the particle's velocity to be consistently opposite to the active noise, which can be realised by having the thermal noise not only opposite to but also larger in magnitude than the active noise. The result is again a trajectory which displays little displacement from its initial position. It is interesting to notice that the latter type of trajectory is only allowed in the presence of thermal noise: letting the thermal diffusion coefficient $D_T$ to vanish would cause the whole branch of $g(\lambda)$ with negative derivative to disappear, signalling that only positive values of the active work are allowed.

\section{Confined Active Ornstein-Uhlenbeck particles}
\label{sec:potential}

We now apply the numerical techniques discussed in \autoref{ssec:cloning} to AOUPs subject to different confining potentials. For each case, we present a numerical estimate of the SCGF obtained with a cloning algorithm and compute the rate function as a  Legendre-Fenchel transform. We then compare such rate functions with those estimated directly from the logarithm of the empirical distribution of the active work. In addition, by looking at biased trajectories, we explore the mechanisms leading to active work fluctuations in the various potentials considered.

It is worth recalling the phenomenology of the fluctuations of the injected power for a confined Brownian particle, as it provides a reference frame for a better understanding of the confined AOUP problem. The confined Brownian problem was first examined in~\cite{farago}, where the author concludes that an external potential has only pre-asymptotic effects on the fluctuations of the injected power, therefore the rate function remains equal to that of the free-particle problem. In simple terms, the asymptotic fluctuations of the injected power are not affected by the presence of a confining potential. This picture is consistent with the intuitive explanation of the general shape of the rate function, which is caused by the positive feedback between fluctuations of the noise and energy uptake of the particle mentioned at the end of~\autoref{ssec:saddle-point}. Much more recent than its passive counterpart, the problem of a confined active particle is currently under the scrutiny of the active matter community~\cite{aoup1, activeconfined1,activeconfined2, majum1,activeconfined4}. It is well understood, for instance, that the two timescales of the problem---one related to the confining potential and the other to the persistent noise---compete in the determination of the steady-state distribution of the particle's position. Thus, depending on the parameters of the problem, in steady state the active particle will be either pushing against the slope of the confining well or fluctuating in the middle of the well in the low persistence limit.

Within the specific context of the AOUP model, it is interesting to notice the singularity of the harmonic case with respect to generic confining potentials. In this case, as in the free problem, the dynamics satisfies detailed balance~\cite{szamelAOUP,cengio2020fluctuation}, resulting in an {\it effective equilibrium regime}. As a result, the harmonically confined AOUP does not display the `pushing' phase in steady state for any value of the parameters. Such peculiar aspect of AOUPs is reflected in the fluctuations of the active work. In fact, our study reveals and highlights the differences in the phenomenology of active work fluctuations between the harmonic case (\autoref{ssec:harmonic}) and cases where the particle is confined by nonlinear confining potentials (\autoref{ssec:nonlinear}), at variance with the passive problem where the confining potential has little-to-no effect~\cite{farago}.

\subsection{Harmonic potential}\label{ssec:harmonic}

The AOUP of this section interacts with a confining harmonic potential $U(\bm{r})\,{=}\,k \r^2 /2$, so that the equations of motion in the overdamped limit are
\begin{subequations}\label{eq:AOUP-harmonic}
 \begin{align}
  \label{eq:AOUP-position_harm} \dot{\bm{r}}(t) &= \bm{v}(t) -\gamma^{-1}k \bm{r}(t) + \sqrt{2 D_T}~ \bm{\xi}(t),\\
  \label{eq:AOUP-propulsion_harm} \dot{\bm{v}}(t) &= -\gamma_R \bm{v}(t) + \sqrt{2 D'_R}~ \bm{\eta}(t).
 \end{align}
\end{subequations}
As mentioned in the previous paragraph, the harmonic potential is somewhat special for the AOUP, as it results in linear equations of motion. Therefore, the position and active force processes $\bm{r}(t)$ and $\bm{v}(t)$ are Gaussian, like the noises $\bm{\xi}(t)$ and $\bm{\eta}(t)$. Furthermore, the problem can be shown to have an effective formulation as an equilibrium problem~\cite{szamelAOUP} and the steady-state distribution of the AOUP position is a Gaussian distribution centered at the bottom of the potential well. From the perspective of active work fluctuations, there does not seem to be a fundamental difference with respect to the free case of~\autoref{sec:free}, apart from an obvious reduction of the average active work. In fact, after solving~\autoref{eq:AOUP-position_harm} for $\bm{r}(t)$, it is straightforward to show that (assuming vanishing position and self-propulsion speed at $t\,{=}\,0$ for simplicity)
\begin{equation}\label{eq:AOUP-harmonic-mean-value}
 \avg{\bm{v}(t)\cdot\dot{\bm{r}}(t)} = \avg{\bm{v}(t)\cdot\bm{v}(t)} - \gamma^{-1} k \int_0^t dt'\,\avg{\bm{v}(t)\cdot\bm{v}(t')} \xrightarrow{t\to\infty} \frac{dD'_R}{\gamma_R} \frac{\gamma_R}{\gamma_R + \gamma^{-1}k}.
\end{equation}
The average active work (per unit time) is obtained by multiplying the expression above by $\gamma$: as $k\,{>}\,0$, the harmonic average is always smaller than the free average $\gamma d D'_R / \gamma_R$.

\begin{figure}[ht!]
\begin{center}
  \begin{tabular}{cc}
       \includegraphics[width=0.5\columnwidth]{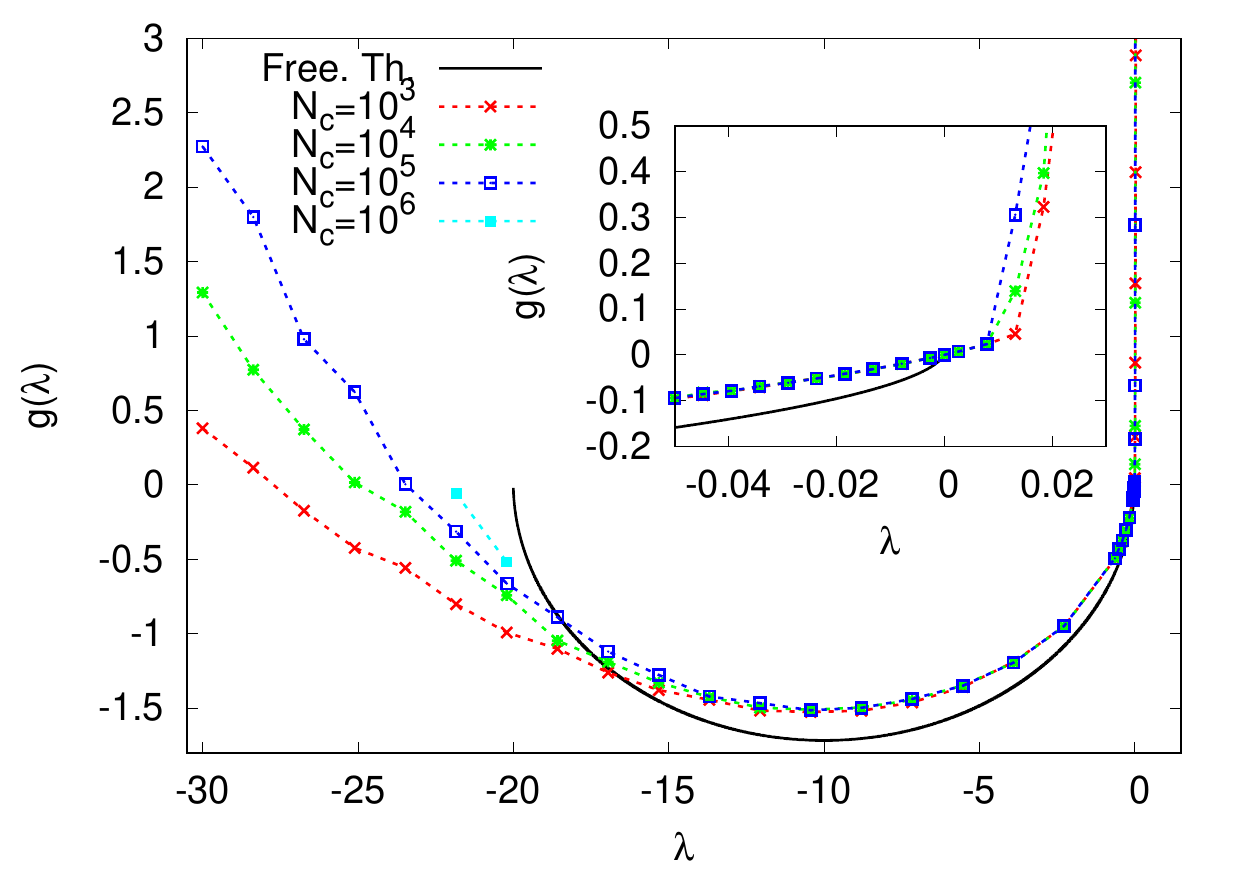}
       \includegraphics[width=0.5\columnwidth]{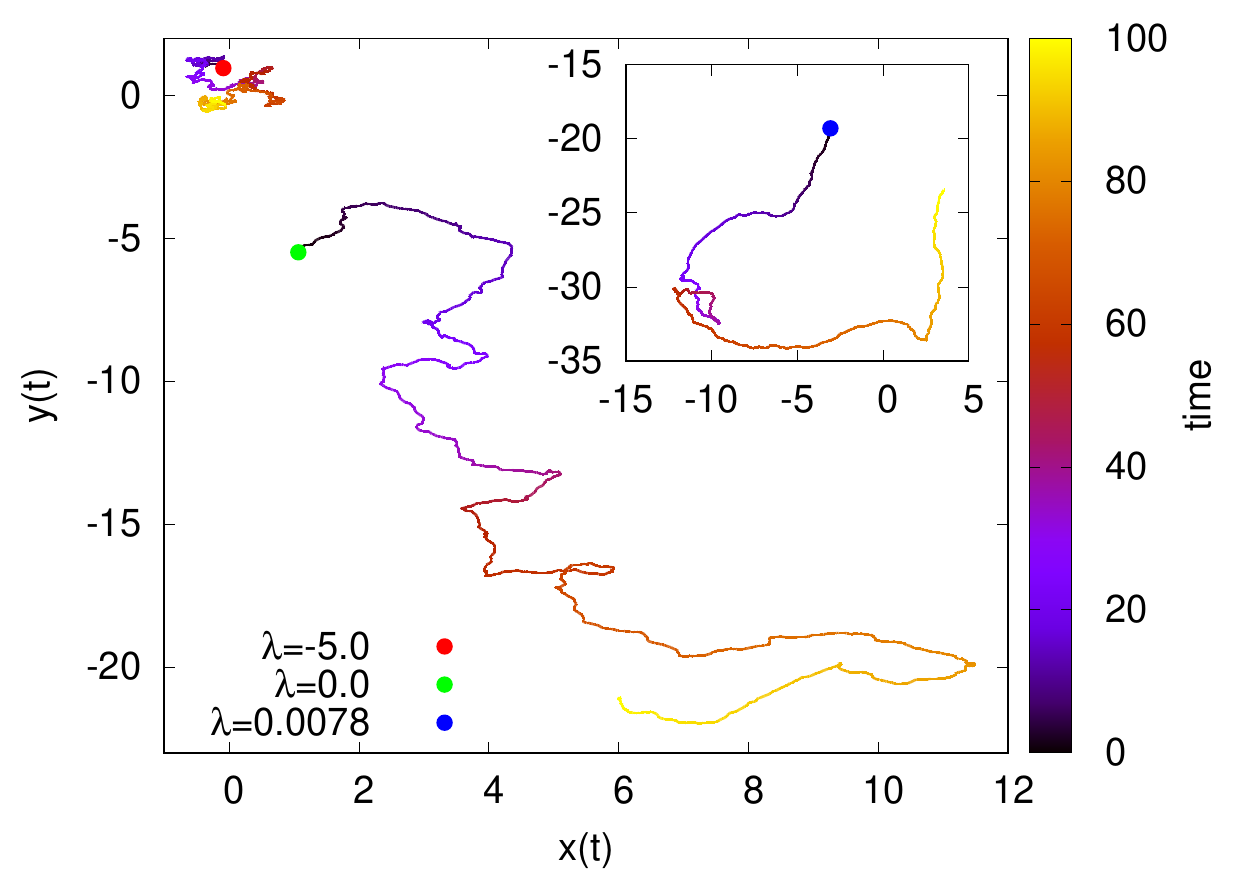}\\
  \end{tabular}
\caption{\footnotesize{Left: SCGF of the AOUP confined by a harmonic potential $k\r^2/2$, with $k\,{=}\,1.0$---the relevant features of the SCGF do not change by varying $k$. The points are obtained via the cloning algorithm and curves obtained with a different number of clones $N_c$ are compared. The analytical SCGF of the free problem is also shown for comparison as a black solid line (Free Th. in the key). The inset shows a zoom of the main figure around $\lambda=0$. Although the maximum number of clones used for this work did not allow convergence of the leftmost part of the curve, our results indicate that the SCGF of the harmonic problem might still be defined for all $\lambda\,{<}\lambda_2^{\text{harm}}$. Right: sample trajectories of the clones corresponding to three representative values of $\lambda$--positive, negative and zero (starting points respectively represented as blue, red and green circles, sampled from the position distribution of the biased process). The phenomenology is also similar to the free-particle problem, with the persistent motion typical of AOUPs disappearing as $\lambda$ decreases.}}
\label{fig:harmonic-SCGF-cloning}
\end{center}
\end{figure}

\begin{figure}[ht!]
\begin{center}
  \begin{tabular}{cc}
       \includegraphics[width=0.5\columnwidth]{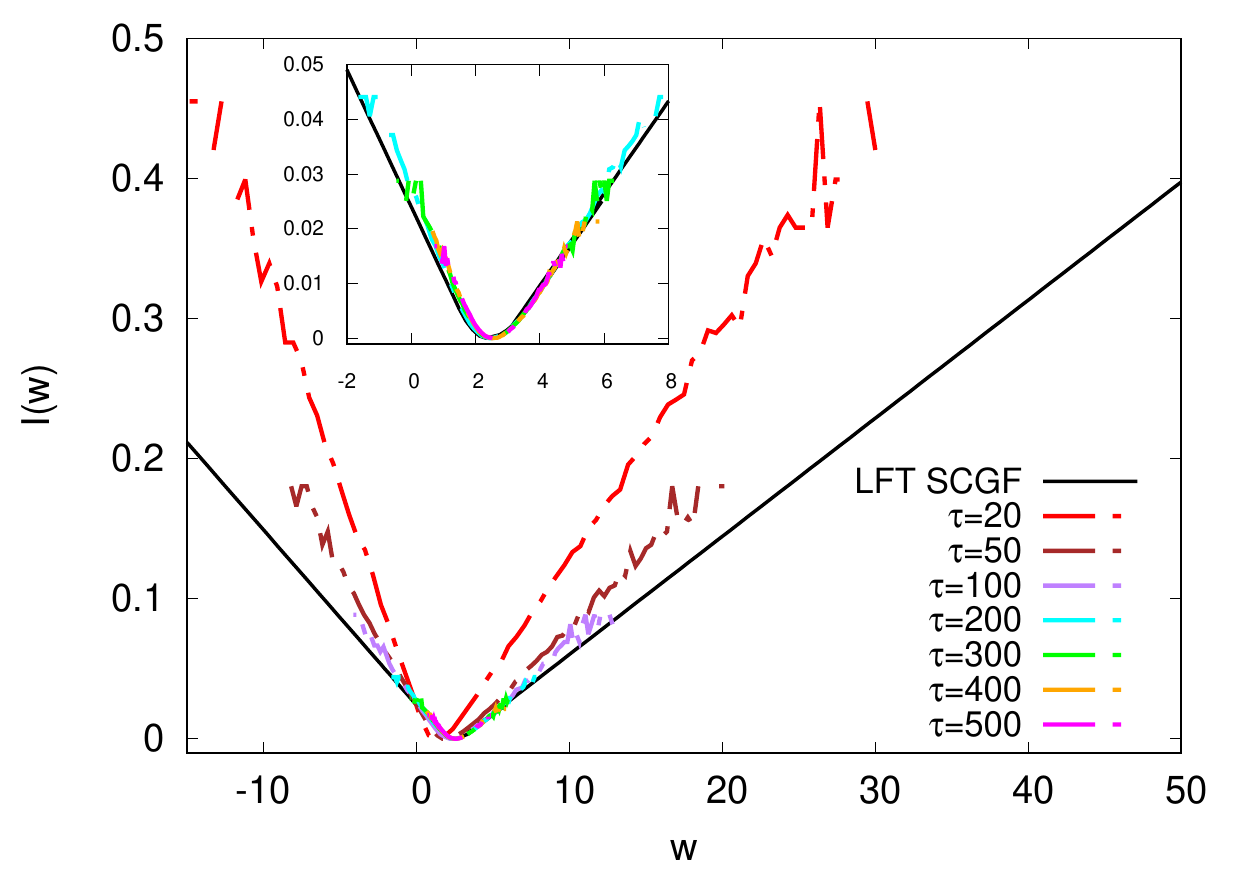}
       \includegraphics[width=0.5\columnwidth]{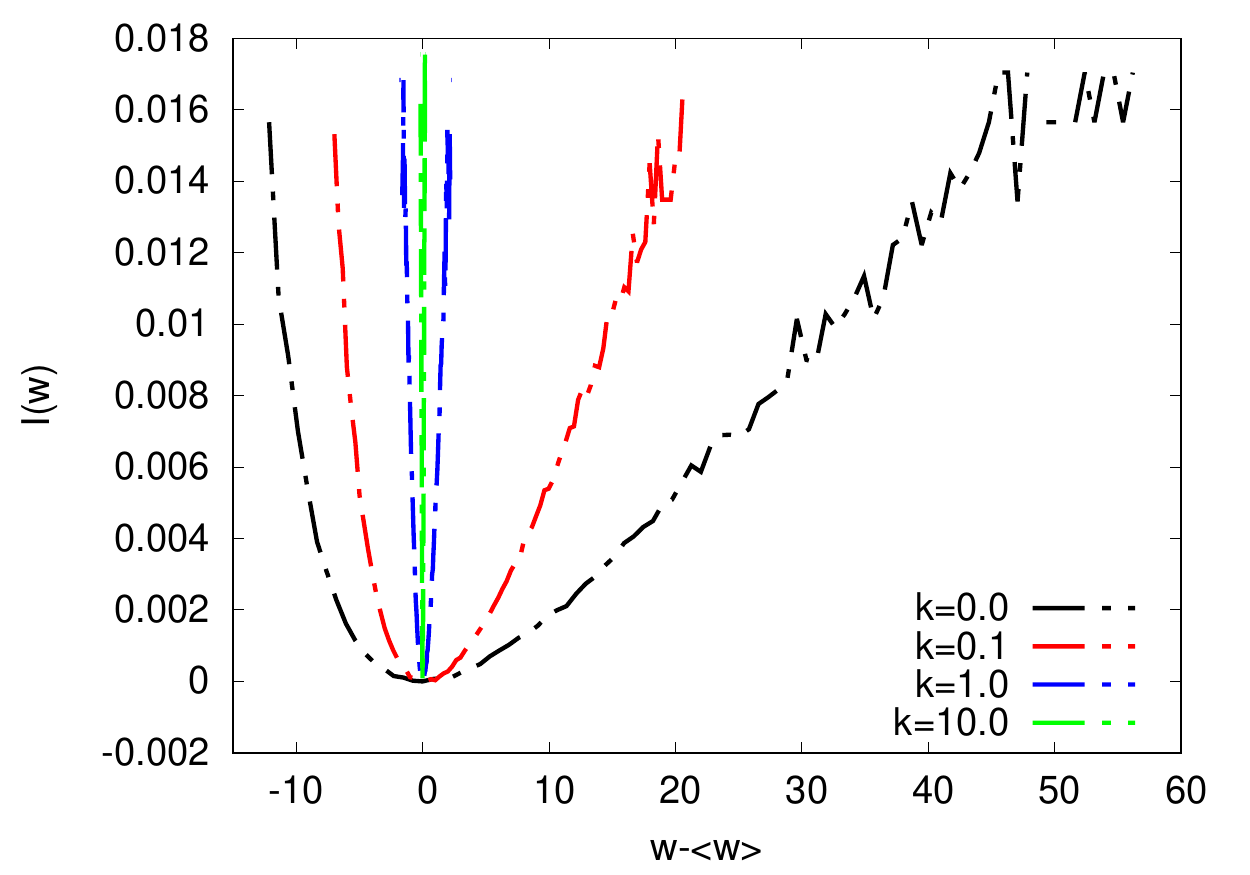}\\
  \end{tabular}
\caption{\footnotesize{Left: direct numerical estimates of the rate function of the active work for the AOUP in the harmonic potential. The parameters are the same as in~\autoref{fig:harmonic-SCGF-cloning}. The figure reports also the Legendre-Fenchel transform of the SCGF obtained via cloning (LFT SCGF in the key). The inset shows instead an enlargement of the main figure around the rate functions minima. The rate function displays linear branches on both sides of the minimum. The cloning estimate of the SCGF suggests that the right branch might be truly linear rather than only asymptotically linear as in the free-particle problem.  Right: Comparison of rate functions translated horizontally with respect to the mean value at $\tau=500$ for various $k$'s in the range $[0, 10]$ as reported in the key. Concerning the rate function, no qualitative changes are observed by varying $k$, even if the rate functions' tails show an increase in their slope with increasing $k$, sign of the stronger confining action of harmonic potentials with higher $k$.}}
 \label{fig:harmonic-rate-function}
\end{center}
\end{figure}

The cloning estimate of the SCGF $g^{\text{harm}}(\lambda)$ is shown in~\autoref{fig:harmonic-SCGF-cloning} (left panel), together with some representative trajectories (right panel). For the sake of comparison, the analytical SCGF of the free case \autoref{eq:SCGF} is also shown in the left panel (black solid line). % to show that the two SCGFs might have a similar functional form, even if definition interval and curvature different.}
The rate function resulting from the LFT of $g^{\text{harm}}(\lambda)$ is shown in~\autoref{fig:harmonic-rate-function} and compared to the direct numerical estimate. The trajectory starting from the green dot in the right panel of~\autoref{fig:harmonic-SCGF-cloning} is an example of typical trajectory ($\lambda\,{=}\,0$). The figure also reports a trajectory with larger-than-average active work ($\lambda\,{>}\,0$, starting from the blue dot in the inset) and one with smaller-than-average active work ($\lambda\,{<}\,0$, red dot). On a qualitative level, the trajectory with a positive bias only looks more persistent than the average trajectory, whereas the one with a negative bias remains closer to the initial condition $\bm{r}_0=0$, as it was the case with negatively biased trajectories of the free problem. However, the cloning estimate of the harmonic SCGF reveals two important differences with respect to the free problem, where the SCGF $g^{\text{free}}(\lambda)$ is defined in an interval $\left[\lambda_1^{\text{free}}, \lambda_2^{\text{free}}\right]$ and it is {\it steep}, i.e. the derivative diverges at the boundaries of the domain. On the one hand, for large positive $\lambda$'s the algorithm returns diverging results, supporting the hypothesis that also the harmonic SCGF diverges at a certain $\lambda\,{=}\,\lambda_2^{\text{harm}}$. In addition, the derivative of the harmonic SCGF in $0$ is much smaller than in the free case and it does not vary significantly for small positive $\lambda$'s. Therefore, our best estimate does not rule out the possibility of a finite derivative of $g^{\text{harm}}(\lambda)$ at the boundary of the domain. We will comment further on this point later, when discussing rate functions.

On the other hand, for $\lambda$ large and negative, our numerics do not reveal any sign of divergence, although a higher number of clones might be required for estimating the actual values of $g^{\text{harm}}(\lambda)$. In other words, what we observe here is analogous to what observed in the free problem for $\lambda$ slightly bigger than $\lambda^{\text{free}}_1$, where, because of the large magnitude of $\lambda$, the largest number of clones we can afford does not grant the coincidence of numerical estimate and analytical prediction. We conclude that, at variance with the free AOUP case, the SCGF of the active work might not diverge for large negative $\lambda$, thus having a domain of the form $(-\infty, \lambda_2^{\text{harm}}]$.

We have also obtained an independent estimate of the rate function itself, by sampling the active work in direct numerical solutions of~\autoref{eq:AOUP-harmonic}. With respect to higher-than-average fluctuations, the difference between a steep $g(\lambda)$, the derivative of which diverges at the boundary of the domain, and a non-steep one is the following: the rate function associated with a steep SCGF is only asymptotically linear, with the slope of the linear branch coinciding with the domain boundary $\lambda_2$, whereas a non-steep SCGF results in a rate function $I(w)$ which is exactly linear after a threshold $w^*$. The threshold coincides with the limiting slope of the SCGF, i.e. $w^*\,{=}\,g'(\lambda_2)$. The rate function for the harmonically confined AOUP is shown in~\autoref{fig:harmonic-rate-function} and it does, within numerical uncertainty, approach a linear branch right after the minimum. Let us nevertheless stress that our estimate cannot exclude a very steep rise of $g^{\text{harm}}(\lambda)$'s derivative very close to $\lambda^{\text{harm}}_2$, which would result again in a steep SCGF. Also the left branch of the numerical rate functions appears linear and there are no reasons to expect a different phenomenology of lower-than-average fluctuations with respect to the free problem. However, the finiteness of the cloning SCGF indicates that there might be differences very far in the tails, although such regions cannot be accessed by the numerical techniques at hand. The proposed scenario seems to be robust also with respect to variation of $k$, taking into account that as $k$ is increased the average active work decreases (\autoref{eq:AOUP-harmonic-mean-value}) and fluctuations become generally rarer, as shown in the right panel of~\autoref{fig:harmonic-rate-function}.

\subsection{Anharmonic potentials}\label{ssec:nonlinear}

\begin{figure}[ht!]
\begin{center}
  \begin{tabular}{cc}
       \includegraphics[width=0.32\columnwidth]{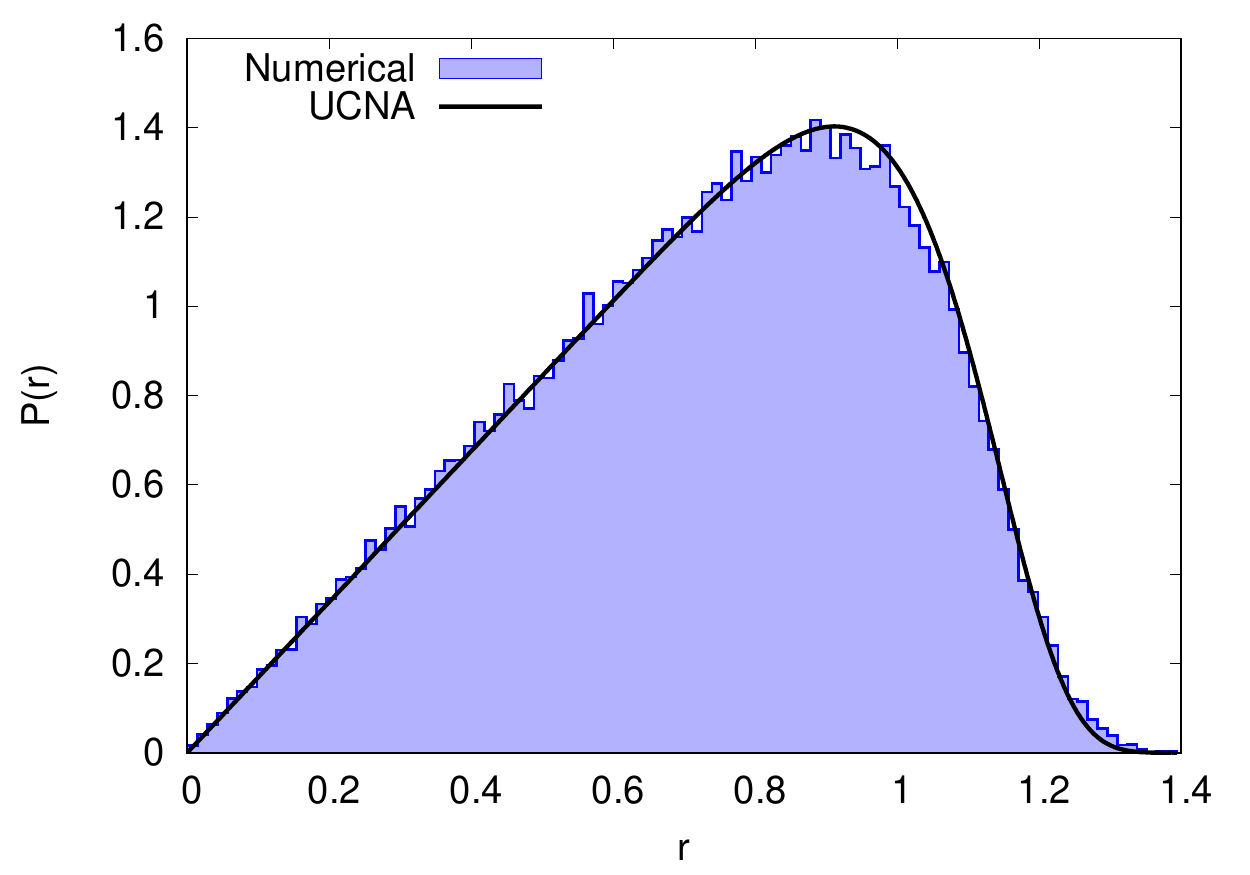}
       \includegraphics[width=0.32\columnwidth]{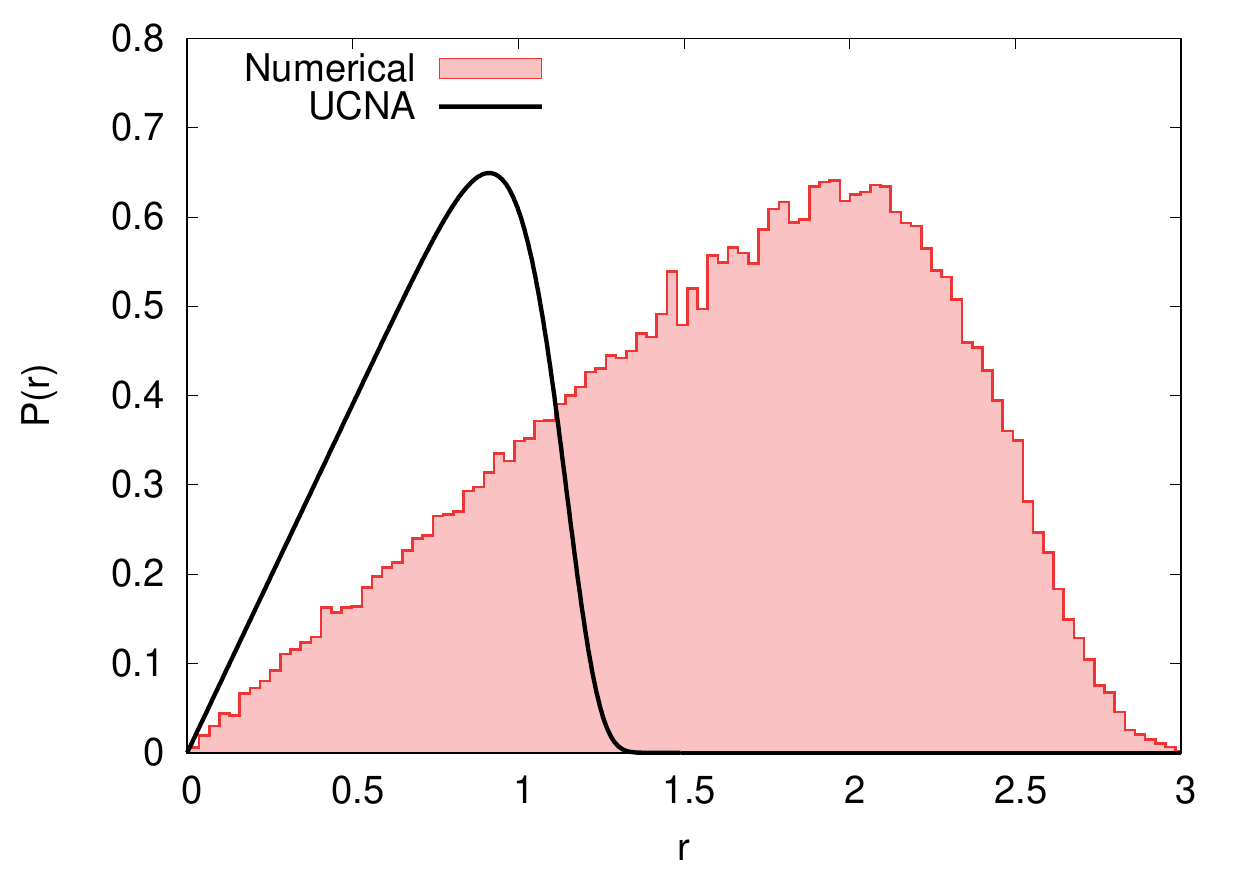}
       \includegraphics[width=0.32\columnwidth]{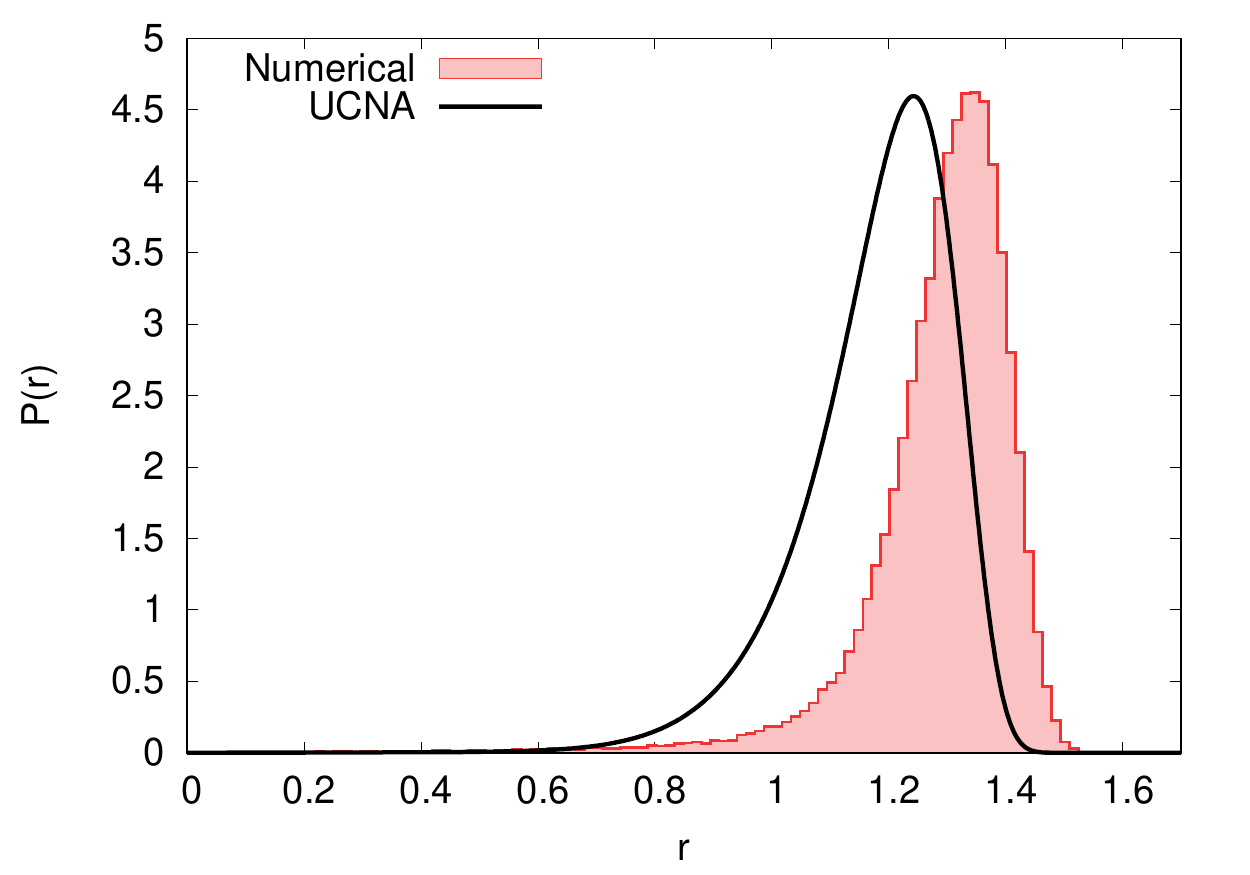}
  \end{tabular}
\caption{\footnotesize{Steady-state distribution of the distance from the center of the well for a single AOUP in the 'stiff' potential $U_{stiff}(\r)=k_{stiff}\r^{10}/10$, both from numerical solutions of the equations of motion and from a Unified Coloured Noise Approximation (UCNA) as in~\cite{aoup1}. Left: $T\,{=}10^2$ in absence of thermal noise $\bm\xi(t)$. Here the UCNA estimate is in good agreement with the numerical distribution. Center:  $T\,{=}10^2$ with both thermal and active noise. Here the UCNA result follows only qualitatively the numerical distributions because of the extra thermal noise.  Right: $T\,{=}\,5 \times 10^{-2}$ with thermal noise $\bm\xi(t)$. Even if the agreement is not perfect, the UCNA still gives a qualitative intuition about the numerical distributions, as in the central panel case.}}
\label{fig:steady-state-stiff}
\end{center}
\end{figure}

In this section we let the AOUP interact with two anharmonic, radially-symmetric confining potentials. The dynamics obeys~\autoref{eq:AOUP1-overd}: in the first case we examine the potential is $U_{\text{stiff}}(\bm{r})\,{=}\,k_{\text{stiff}} \bm{r}^{10}/10$, which we refer to as the `stiff' potential. In such confining potential the AOUP displays a different steady-state behaviour depending on the P\'eclet number $Pe$. Three representative cases are shown in ~\autoref{fig:steady-state-stiff}, where all the parameters but $k_BT$ are fixed, and we consider the presence and also the absence of the thermal noise $\bm{\xi}(t)$. For high temperatures and low P\'eclet, on the left and center, the steady-state distribution of the AOUP position is peaked at the origin, whereas for low temperatures and high P\'eclet, on the right, it displays an annular peak with a finite radius. We study the large deviations of the active work in the regime of parameters resulting in an annular steady-state distribution for the position in presence of the thermal noise, as in the right panel of~\autoref{fig:steady-state-stiff}, and report here the corresponding results. In the other regime, where the steady-state distribution of the position is peaked at the origin, one would only observe progressively symmetric rate functions, as the thermal contribution to active work fluctuations begins to dominate over the active contribution.

\begin{figure}[ht!]
\begin{center}
  \begin{tabular}{cc}
       \includegraphics[width=0.5\columnwidth]{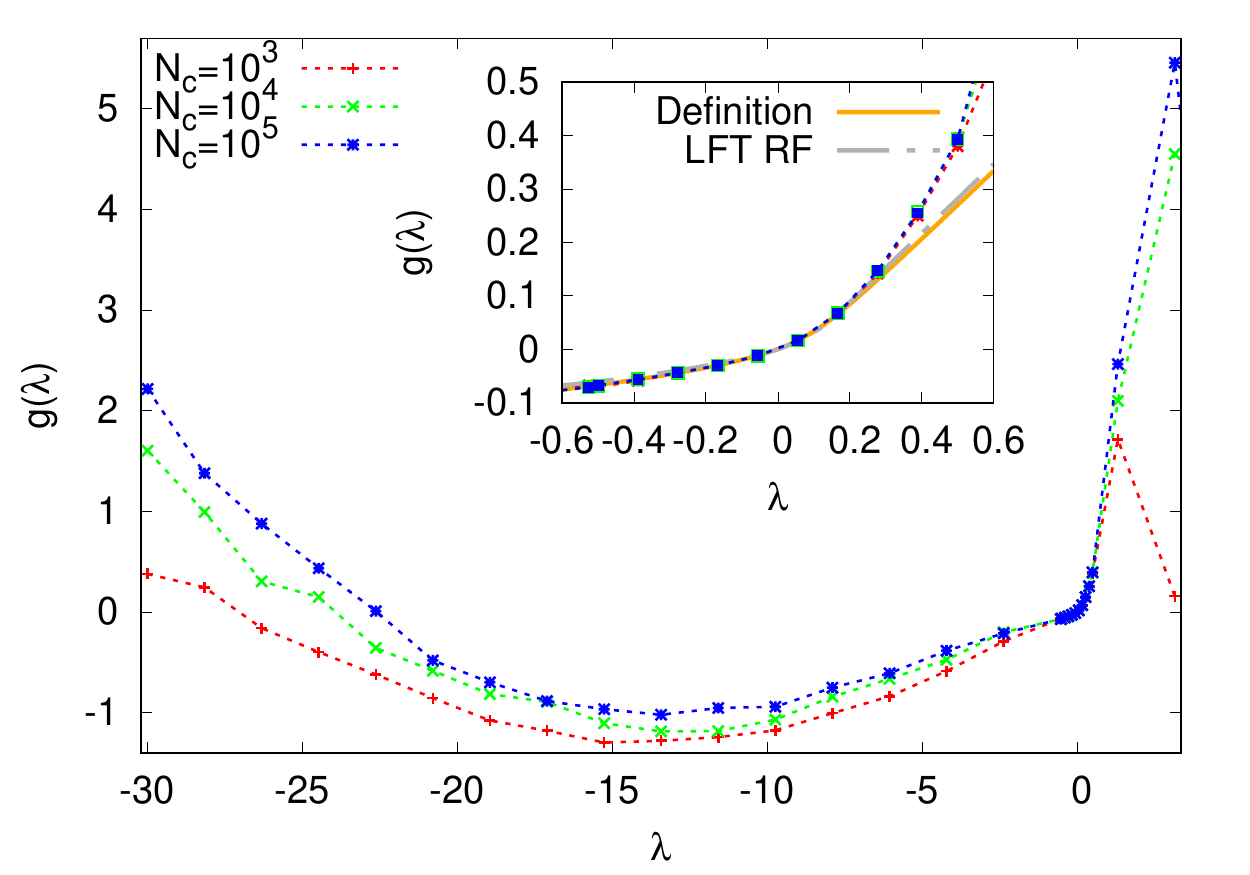}
       \includegraphics[width=0.5\columnwidth]{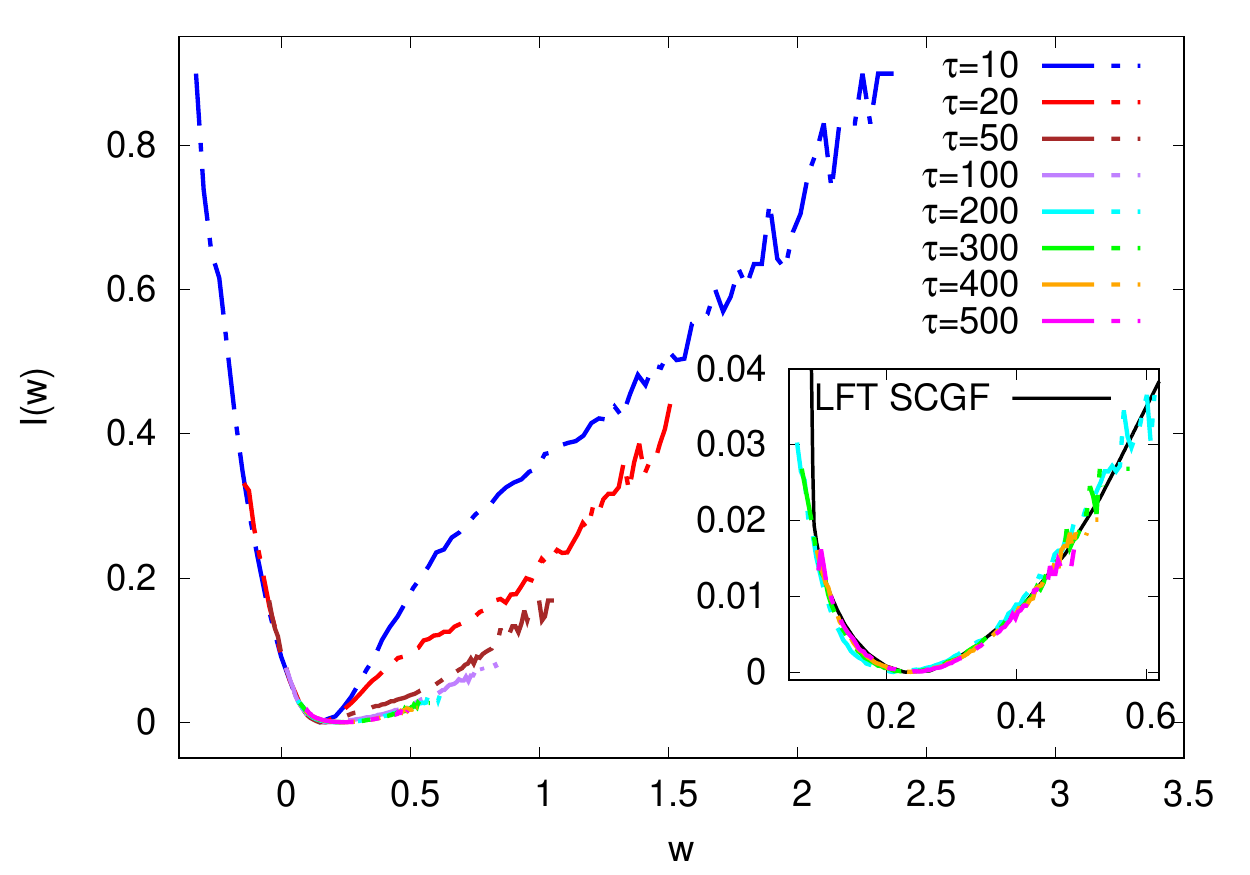}\\
  \end{tabular}
\caption{\footnotesize{Left: SCGF of the AOUP confined by the 'stiff'  potential $U_{stiff}(\r)=k_{stiff} \bm{r}^{10} / 10$, with $k_{stiff}\,{=}\,1.0$. As before, the estimates obtained with a different number of clones $N_c$ are compared. Other parameters are $\gamma\,{=}\,10$, $F_a\,{=}\,20$ and $k_BT\,{=}\,0.05$ (the label `Definition' in the inset key denotes the SCGF evaluated as in~\autoref{eq:SCGF_def} whereas `LFT RF' denotes the Legendre-Fenchel Transform of the numerical Rate Function at the maximum observation time available). Right: estimate of the rate function from direct numerical solutions of the equations of motion (LFT transform of the cloning SCGF is shown as a black solid line in the inset)}}
\label{fig:stiff-SCGF-cloning}
\end{center}
\end{figure}

Also in this case we provide an estimate of the SCGF $g^{\text{stiff}}(\lambda)$ obtained via cloning and an estimate of the rate function obtained by sampling the empirical distribution of the active work over several independent numerical solutions, following the method described in \autoref{ssec:cloning}. The results are shown in~\autoref{fig:stiff-SCGF-cloning}, with the SCGF on the left and rate function on the right. As in the free (\autoref{ssec:cloning}) and harmonic (\autoref{ssec:harmonic}) case, the cloning algorithm returns unphysical results for $\lambda$ larger than a certain threshold $\lambda_2^{\text{stiff}}$, indicating that the domain of $g^{\text{stiff}}(\lambda)$ might have an upper extremum. In contrast with the harmonic case, however, our numerical estimates are compatible with a steep SCGF, i.e. with $g^{\text{stiff}}(\lambda)$ reaching an infinite derivative at $\lambda_2^{\text{stiff}}$. Direct numerical estimates of the rate function indeed do not show any linear branch, supporting the hypothesis of a steep SCGF. For $\lambda$ large and negative the cloning algorithm behaves as in the harmonic problem: there are no divergences but a higher number of clones would be required for a quantitative estimate of the SCGF.

\begin{figure}[ht!]
\begin{center}
  \begin{tabular}{cc}
       \includegraphics[width=0.5\columnwidth]{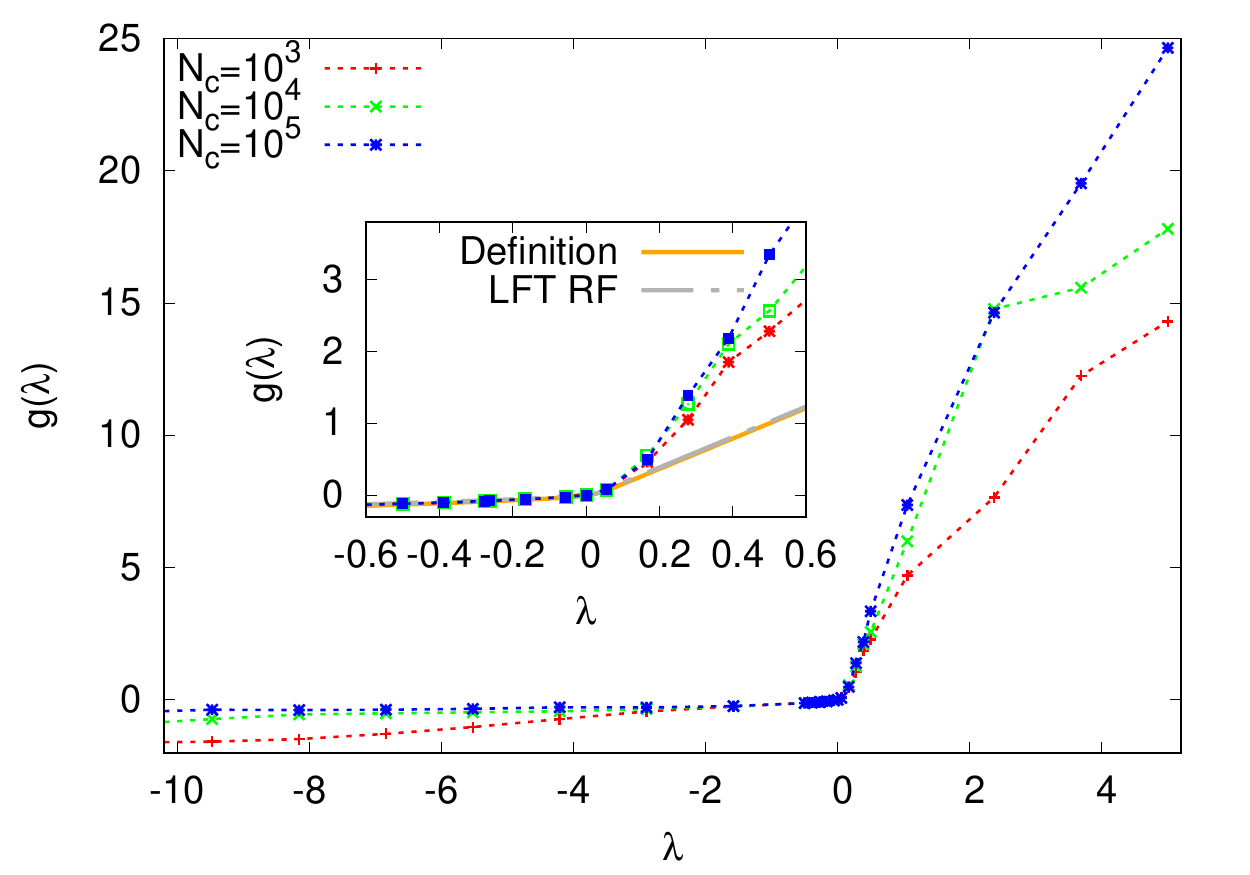}
       \includegraphics[width=0.5\columnwidth]{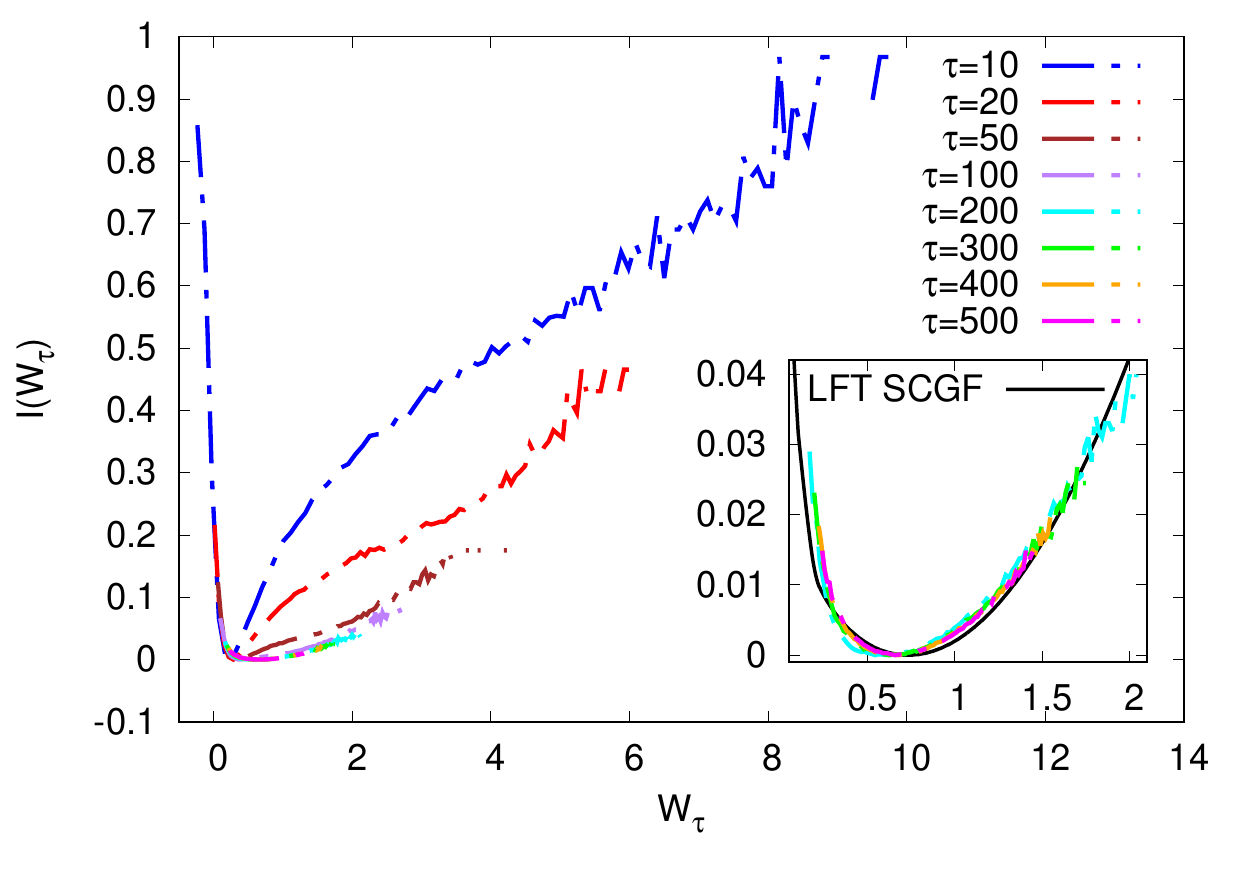}\\
  \end{tabular}
\caption{\footnotesize{Left: SCGF of the AOUP confined in the circular well \autoref{eq:wca_pot} modelled with the WCA potential \autoref{eq:wca_pot2} and with radius with $R=5\sigma$. The parameters of the potential are specified in the main text, the choice of other parameters values is the same as in the previous figures (the labels `Definition' and `LFT SCGF' in the inset key have the same meaning as in the previous figure). Right: direct numerical estimate of the rate function and LFT transform of the SCGF (black solid line).}}
\label{fig:circle-SCGF-cloning}
\end{center}
\end{figure}

We have also studied the large deviations of the active work for an AOUP confined in a circular well, which we have modelled with a Weeks-Chandler-Andersen (WCA) potential on the difference between the distance of the AOUP from the origin $r$ and the radius $R$ of the circular well, i.e.
\begin{equation}
    U_{\text{circle}}(\textbf{r})=\begin{cases}
        0       &\text{for }r<R-2^{\frac{1}{6}}\sigma\\
        U_{LJ}(R-r)-U_{LJ}(2^{1/6}\sigma) &\text{for }r\geq R-2^{1/6}\sigma
        \end{cases}
       \label{eq:wca_pot}
\end{equation}
with $U_{\text{LJ}}(x)$ the Lennard-Jones potential,
\begin{equation}
U_{\text{LJ}}(x) = 4\epsilon\left[\left(\frac{\sigma}{x}\right)^{12}-\left(\frac{\sigma}{x}\right)^{6}\right].
       \label{eq:wca_pot2}
\end{equation}
We set both the spatial ($\sigma$) and energy ($\epsilon$) scales of the Lennard-Jones potential to $1$. In practical terms, the AOUP moving in the potential $U_{\text{circle}}(\bm{r})$ is free until its distance from the origin reaches $R-2^{1/6}$, then it is pushed back by the ascending branch of the Lennard-Jones potential. Also for this potential the steady-state distribution of the AOUP position is either annular or peaked at the origin depending on the P\'eclet number and we focus on the regime where it is annular. Our estimates of SCGF and rate function are shown in~\autoref{fig:circle-SCGF-cloning} for $R\,{=}\,5\sigma$. The features of SCGF and rate function are similar to those observed with the stiff potential, therefore we will not comment any further.


\begin{figure}[ht!]
\begin{center}
  \begin{tabular}{cc}
       \includegraphics[width=0.5\columnwidth]{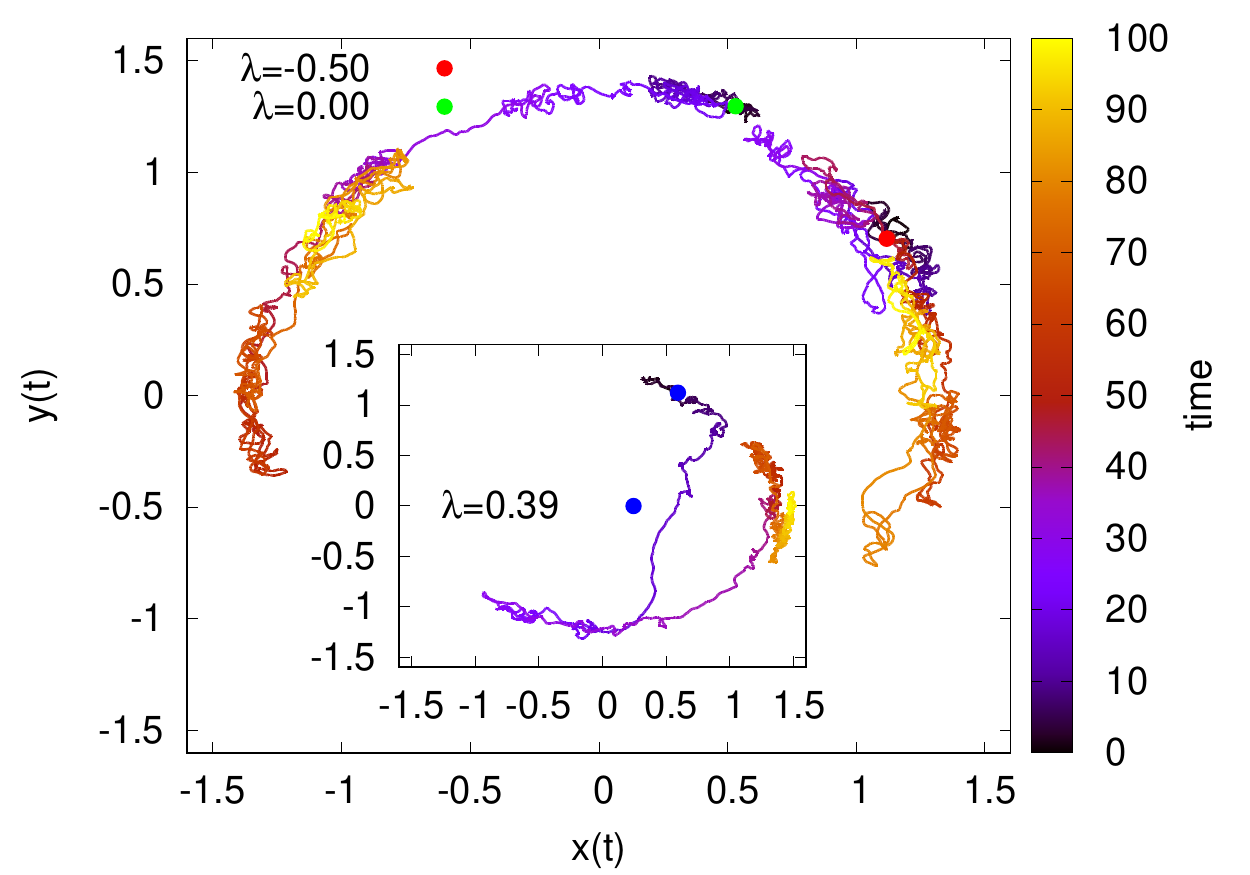}
       \includegraphics[width=0.5\columnwidth]{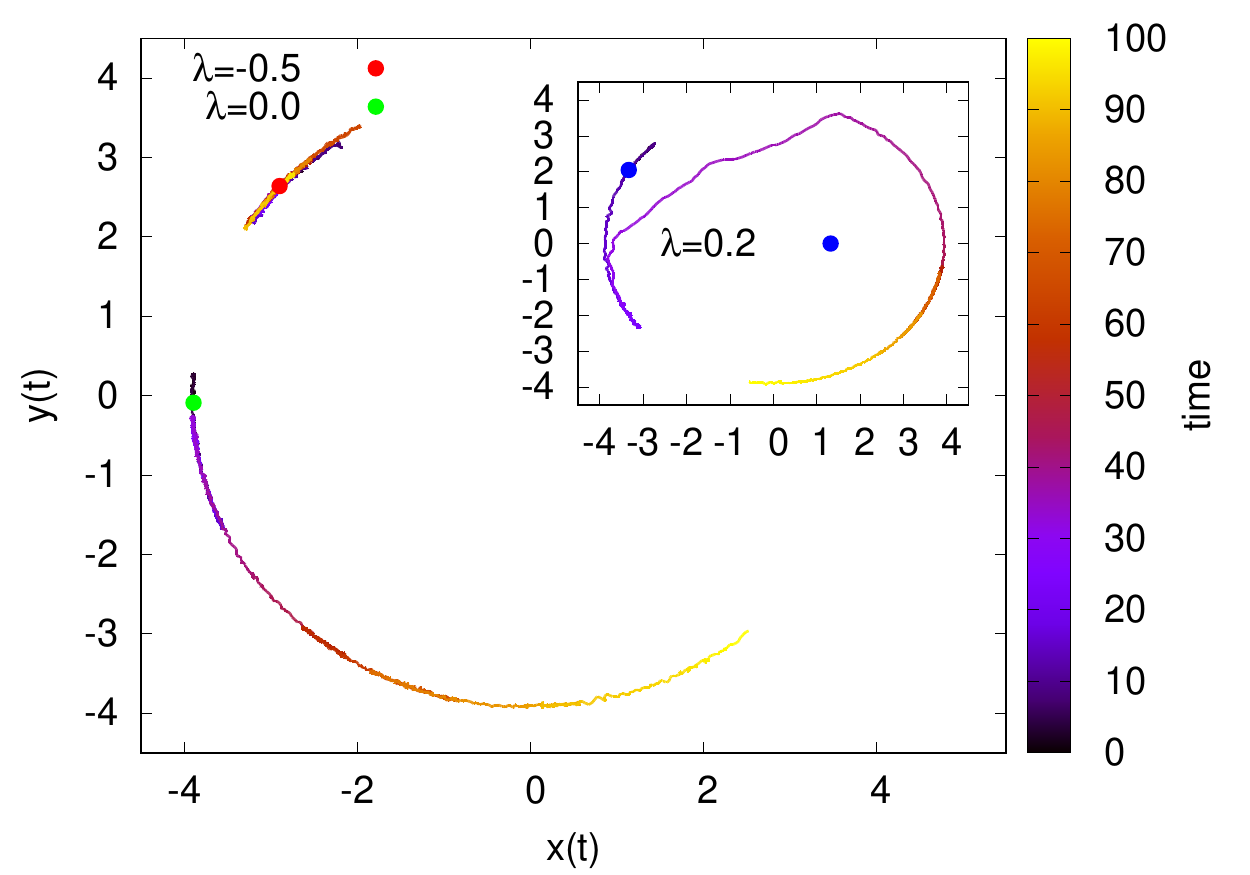}\\
  \end{tabular}
\caption{\footnotesize{Sample trajectories of the cloned trajectories corresponding to three representative values of $\lambda$ --- positive, negative and zero--- in the stiff potential  $U_{stiff}(\r)=k_{stiff} \bm{r}^{10} / 10$, (left) and in the circular well \autoref{eq:wca_pot} (right). The starting points are coloured according to the value of $\lambda$---blue for positive, red for negative and greed for zero---are sampled from the position distribution of the biased process.}}
\label{fig:anharmonic-trajectories}
\end{center}
\end{figure}
Let us close the section by examining the biased trajectories for the anharmonic potentials considered in this section, reported in~\autoref{fig:anharmonic-trajectories} (left panel for the stiff potential, right panel for the circular well). With respect to the harmonic confinement, here the interaction with the potential is crucial for the realisation of both typical and atypical trajectories: this is particularly evident when looking at the biased trajectory of the circular-well problem (right panel of the figure). The typical trajectories (starting from the green dots in the figure) mirror the annular structure of the steady-state distribution of the position. From a dynamical perspective, the AOUP performs a persistent motion along the angular direction, switching between the clockwise and anticlockwise directions. These trajectories still produce a positive active work, although smaller than in the free case. The trajectories with negative $\lambda$ (starting from the red dots in the figure) are those that produce approximately vanishing active work and they do so by having an unusually persistent active noise, so that the particle sits at a distance from the center such that the active force and potential restoring force are balanced while performing thermal fluctuations. By contrast, a larger-than-average active work can be produced when the active noise is less persistent than usual. Having, for instance, a sudden change in the direction of the active noise when the particle is pushing against the potential slope would cause the particle to revert its  direction and slide down that same slope, thus gaining velocity. It is interesting to notice that the mechanisms leading to higher- or lower-than-average active work in the presence of an anharmonic radially symmetric confining potentials are opposite with respect to those at work in the harmonic and free problems: here highly correlated active noise is required to produce a small active work and anti-correlated active noise results in large active work, whereas, in the free problem, unusually correlated active noise results in higher-than-average active work and anti-correlations cause a reduction of the active work.

\section{Active Work Fluctuation Relations}
\label{sec:fl_th}

Fluctuation relations have been formulated as general theorems fixing the symmetry of the distribution of observables $\Omega(t)$ which depend on the system trajectory. These theorems were first proved in the context of non equilibrium diffusion processes, where equilibrium systems are perturbed by external fields, and for jump processes that satisfy time reversal invariance and markovianity \cite{leb_sp, gal_cohen, kurchan_fr, jarz1, ep3}. Later on, fluctuation relations received a more systematic formulation \cite{seifert_fl_th}, and have also been considered in the framework of active matter, using the entropy production as significative thermodynamic observable~\cite{fodorHOWFAR, martinAOUP, catesENTR, capriniAOUP, mandalENTROPY}.
%\footnote{The word {\it detailed} is often used to distinguish such a fluctuation theorem from other forms of fluctuation theorems, as the {\it integral fluctuation theorem} concerning the ensemble mean value of the exponential $e^{-\Omega(t)}$  \cite{seifert_fl_th}.}

A (stationary) {\it detailed fluctuation theorem}, to which we will refer in the following as {\it fluctuation relation}  (FR), is said to be satisfied when the time-independent stationary distribution $\Pi(\omega)$ of the functional observable $\omega=\Omega/\tau$ satisfies the identity 
\begin{equation}
\frac{1}{\tau}\log\left(\frac{\Pi(\omega)}{\Pi(-\omega)}\right)=c~\omega~,
\label{DFT}
\end{equation}
with $c$ a constant depending in general on the system and on the observable considered. When the stationary distribution of $\omega(t)$ satisfies also a large deviation principle, i.e. $\Pi(\omega)\equiv e^{-\tau I(\omega)}$ with $I(\omega)$ rate function, the FR takes the following asymptotic form as $\tau\to\infty$,
\begin{equation}
I(\omega)-I(-\omega)=c~\omega~.
\label{DFTdirect}
\end{equation}
As such, \autoref{DFTdirect} lacks of the subexponential contributions appearing in general also in the stationary configuration distribution and negligible in the large time limit. The validity of an asymptotical FR can actually be traced down to the following property of the SCGF~\cite{leb_sp},
\begin{equation}
g(\lambda)=g(c-\lambda).
\label{eq:scgf_symm}
\end{equation} 

\begin{figure}[t!]
\begin{center}
  \begin{tabular}{cc}
       \includegraphics[width=0.33\columnwidth]{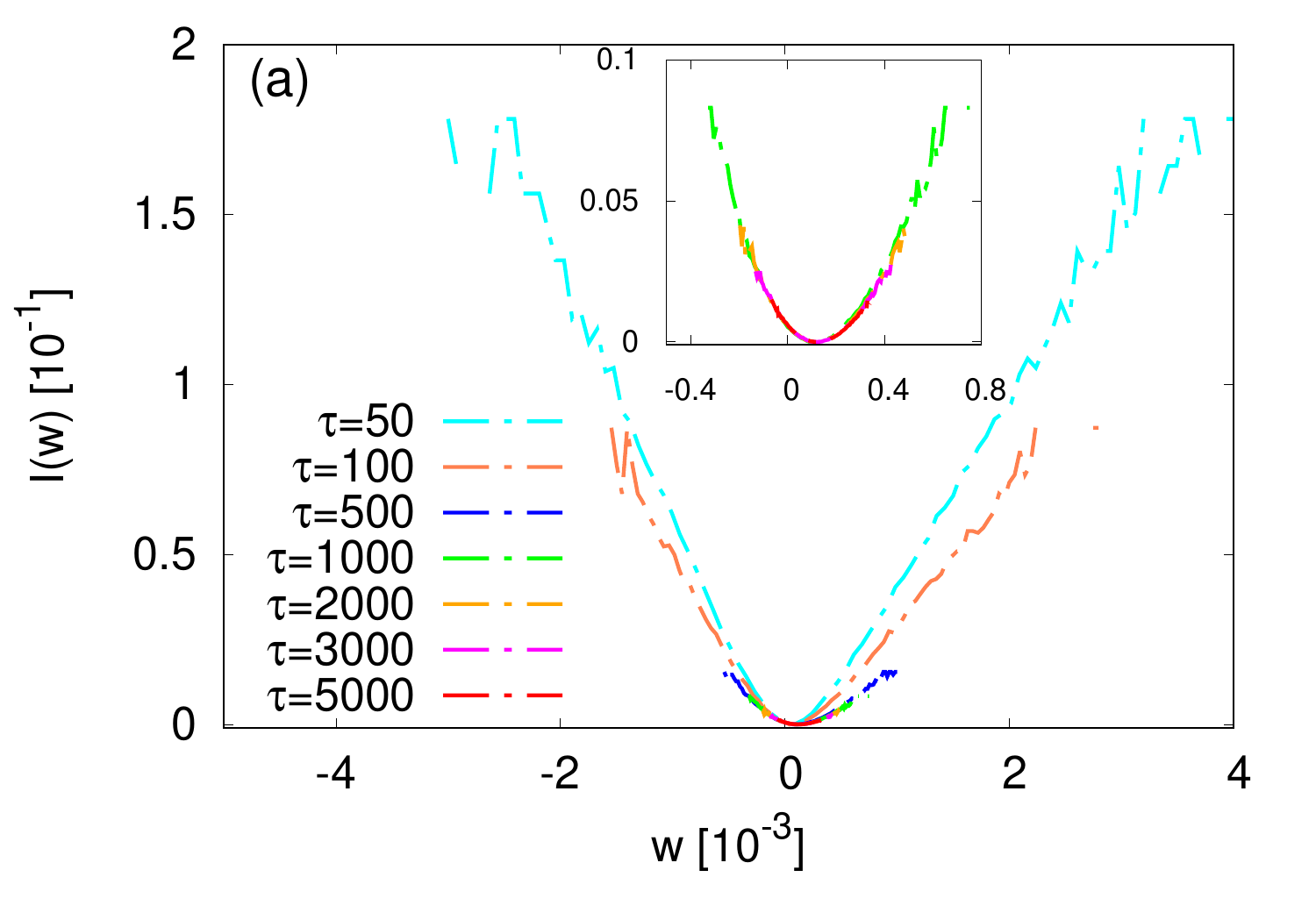}
       \includegraphics[width=0.33\columnwidth]{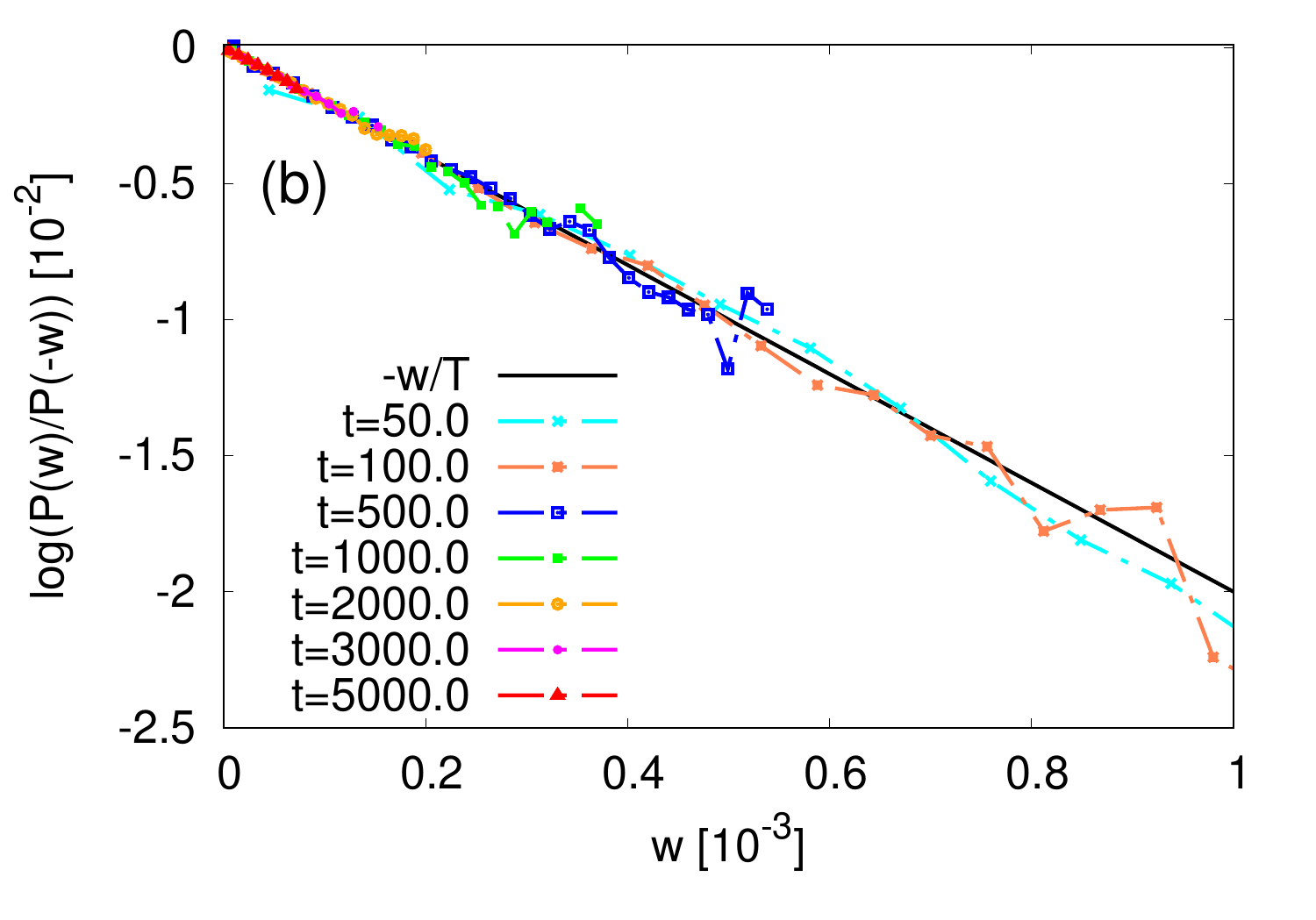}
       \includegraphics[width=0.33\columnwidth]{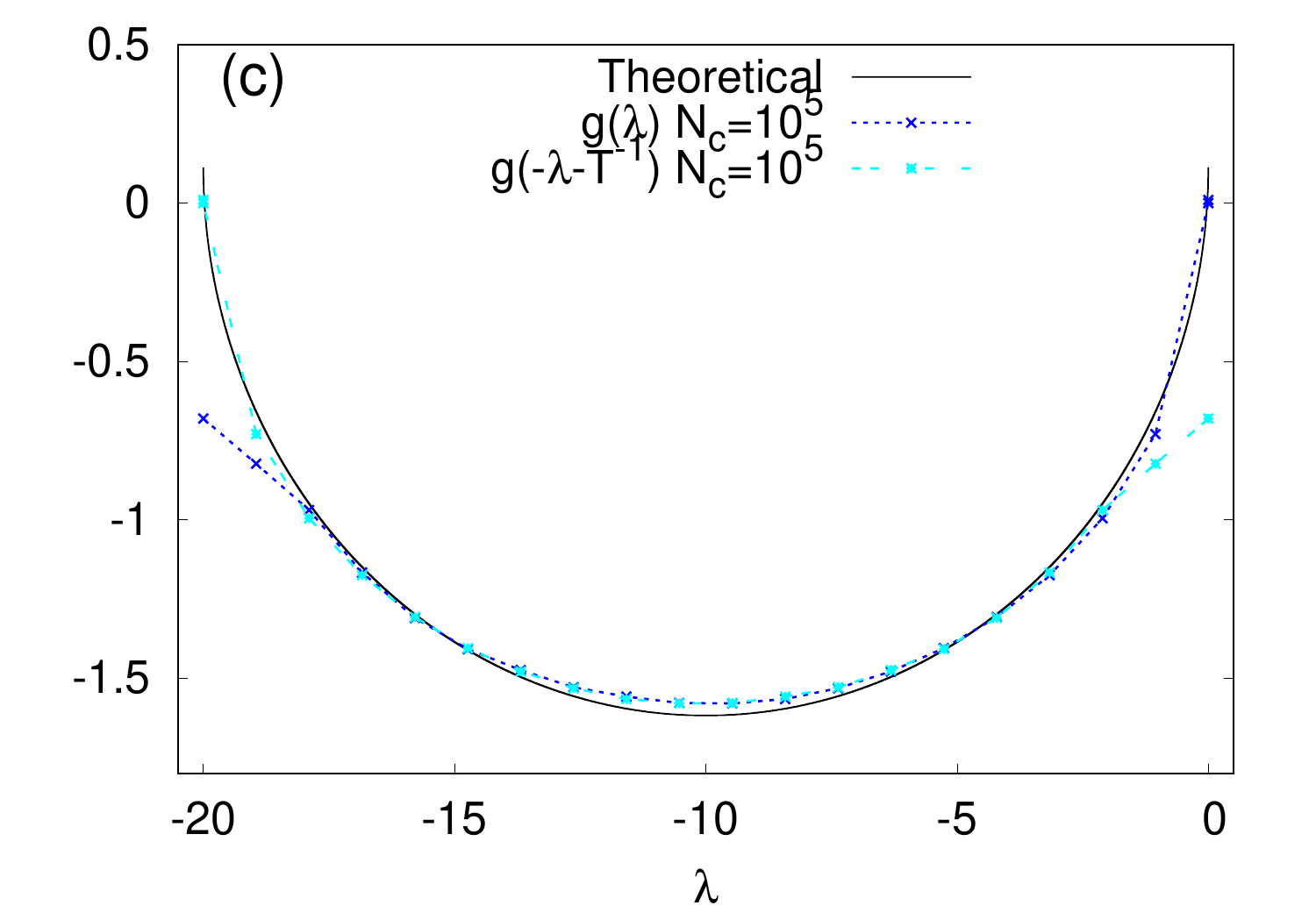}\\
       \includegraphics[width=0.33\columnwidth]{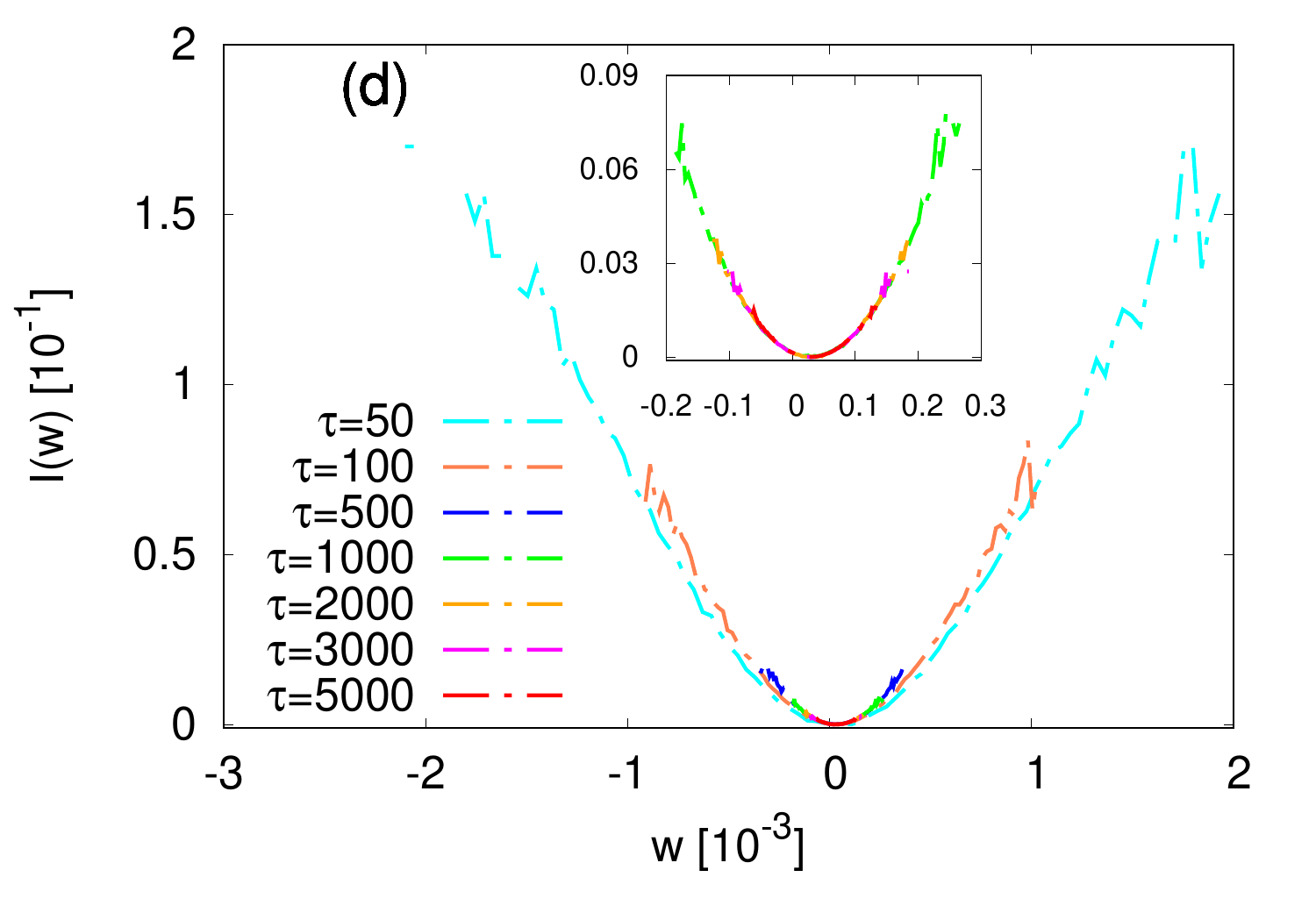}
       \includegraphics[width=0.33\columnwidth]{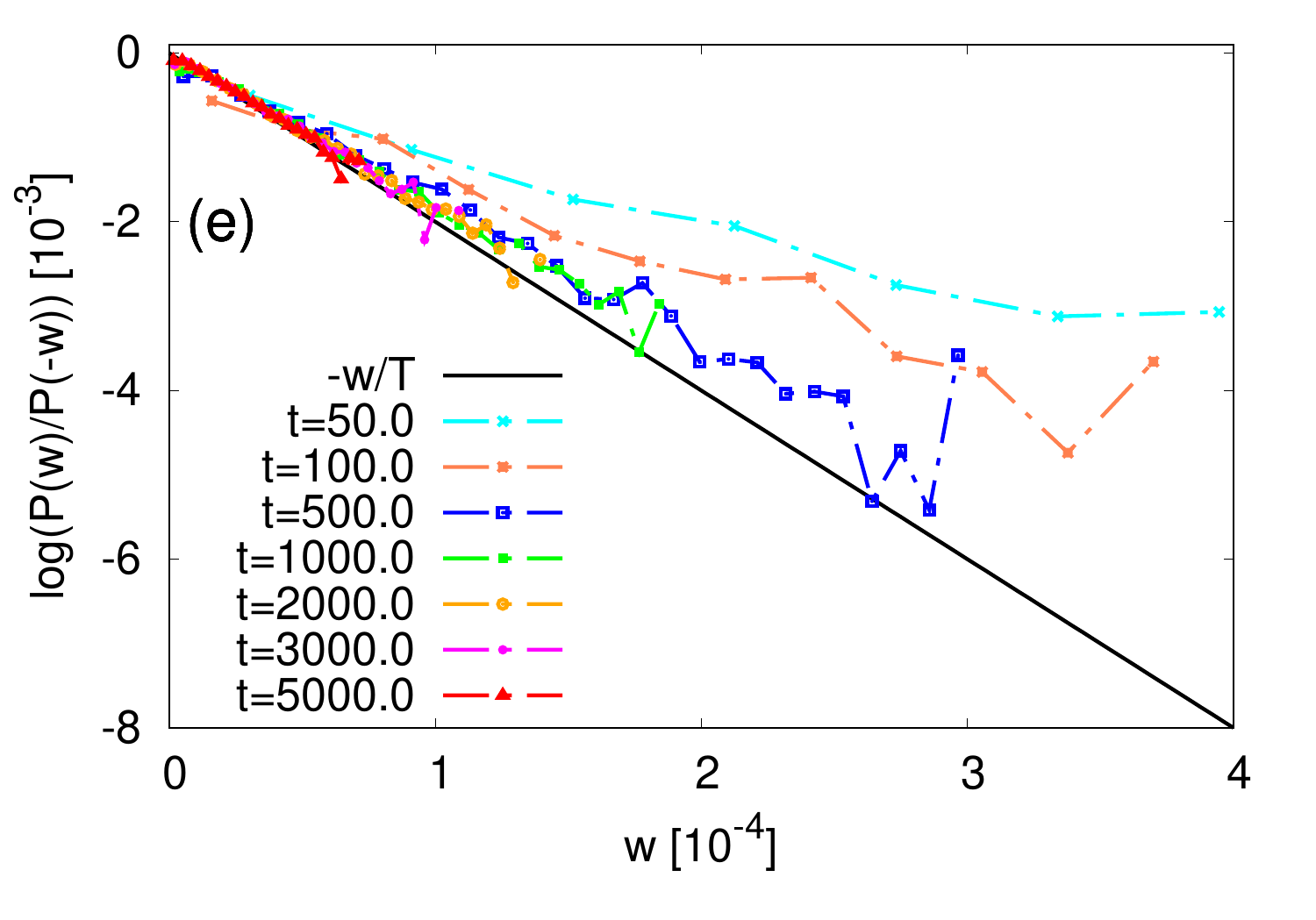}
       \includegraphics[width=0.33\columnwidth]{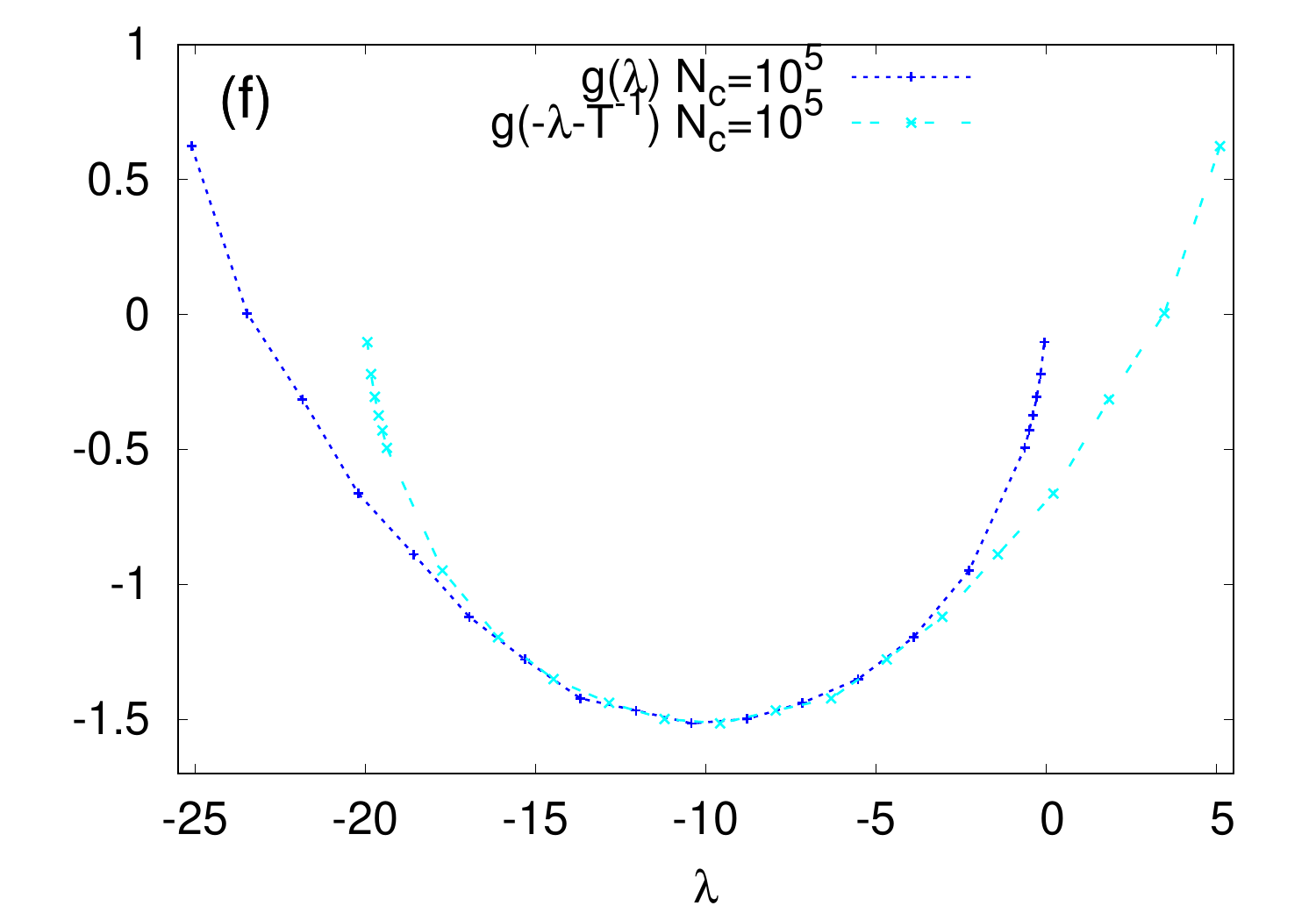}
  \end{tabular}
\caption{\footnotesize{$(a)$ and $(d)$: direct numerical estimates of the rate function of the active work obtained via numerical integration of \autoref{eq:AOUP1-overd} in $d=2$ respectively $(a)$  for a free AOUP with $\gamma=10$, $F_a=0.5$, $k_BT=0.05$ and $(d)$ for an AOUP confined by the harmonic potential $k\bm{r}^2/2$ with $k=1.0, \gamma=10$, $F_a=1.0$, $k_BT=0.05$. The two insets show an enlargement of the main figures around the minima of the rate functions extracted at larger times. $(b)$ and $(e)$: plot of the rate function difference $I(w)-I(-w)$ corresponding respectively to the rate functions in $(a)$ and $(d)$. The black continuous line is $-w/T$.  Note how, especially in the harmonically confined case, the FR reaches a stationary form only after waiting a few persistence times $\gamma_R^{-1}$, when the sub-exponential contributions become negligible. $(c)$: SCGF of the free AOUP in $d=2$ for $T=0.05$, $F_a=20.0$, $\gamma=10$. The figure reports $g(\lambda)$ in the case $N_c=10^5$ taken from \autoref{fig:free-SCGF-rate-cloning}, the symmetric plot $g(-\lambda-T^{-1})$ and the analytical result \autoref{eq:SCGF} (black solid line). $(f)$: SCGF of the AOUP in a harmonic potential in $d=2$ for $T=0.05$, $F_a=20.0$, $\gamma=10$. The figure reports $g(\lambda)$ in the case $N_c=10^5$ taken from \autoref{fig:harmonic-SCGF-cloning} and the symmetric plot $g(-\lambda-T^{-1})$.}}
\label{fig:fr-free-harmonic}
\end{center}
\end{figure}

\begin{figure}[t!]
\begin{center}
  \begin{tabular}{cc}
       \includegraphics[width=0.45\columnwidth]{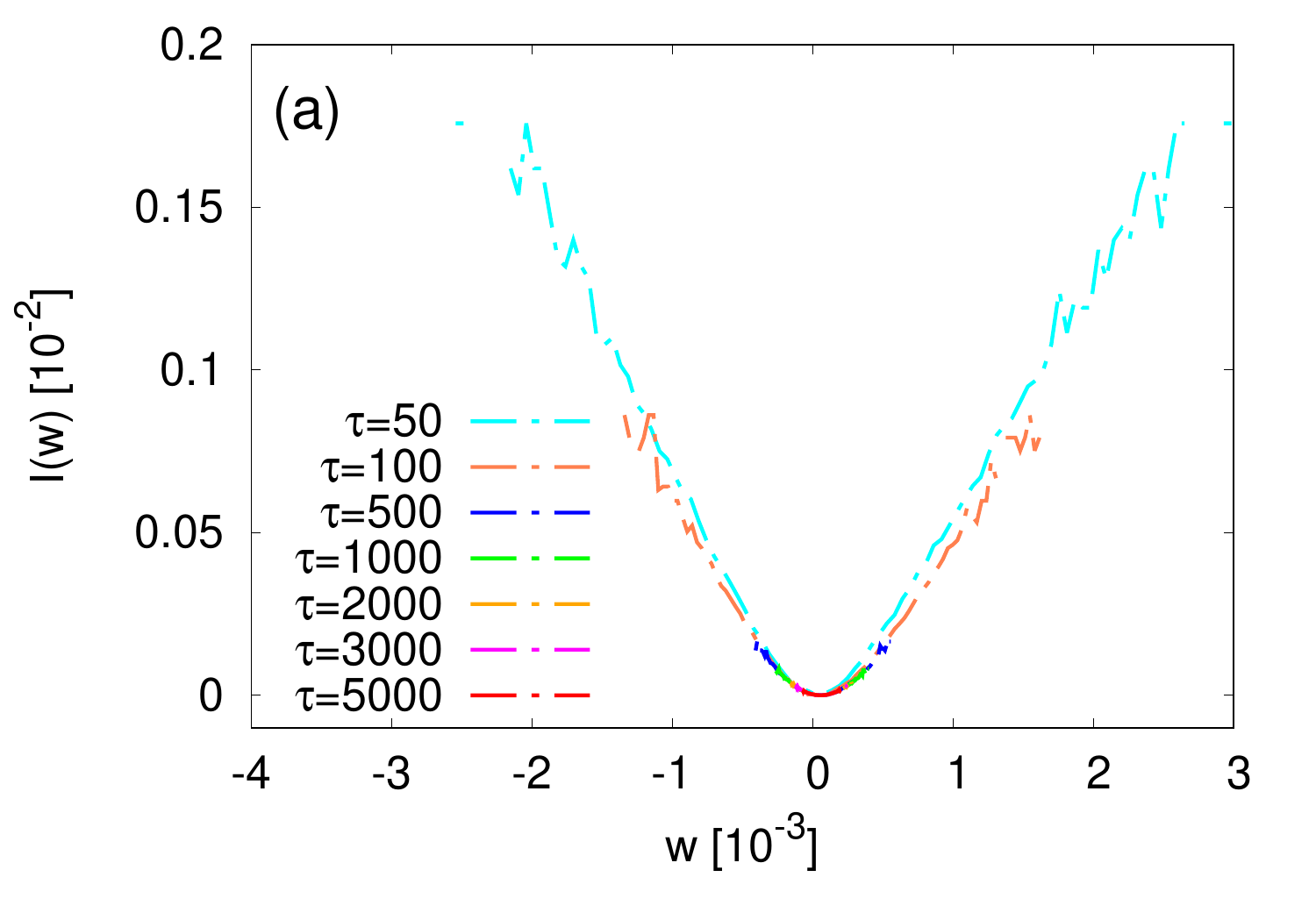}\qquad
       \includegraphics[width=0.45\columnwidth]{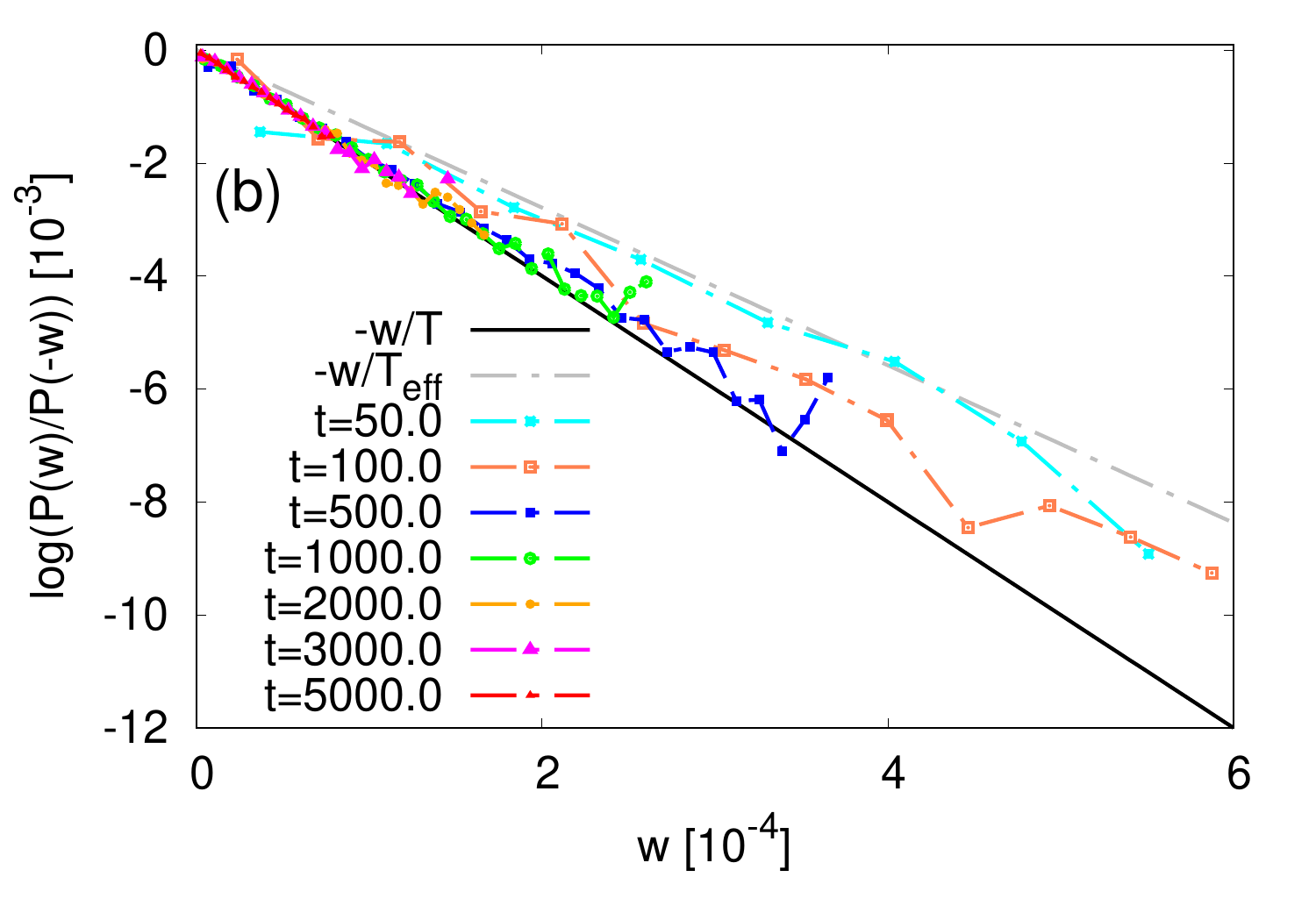}\\
       \includegraphics[width=0.45\columnwidth]{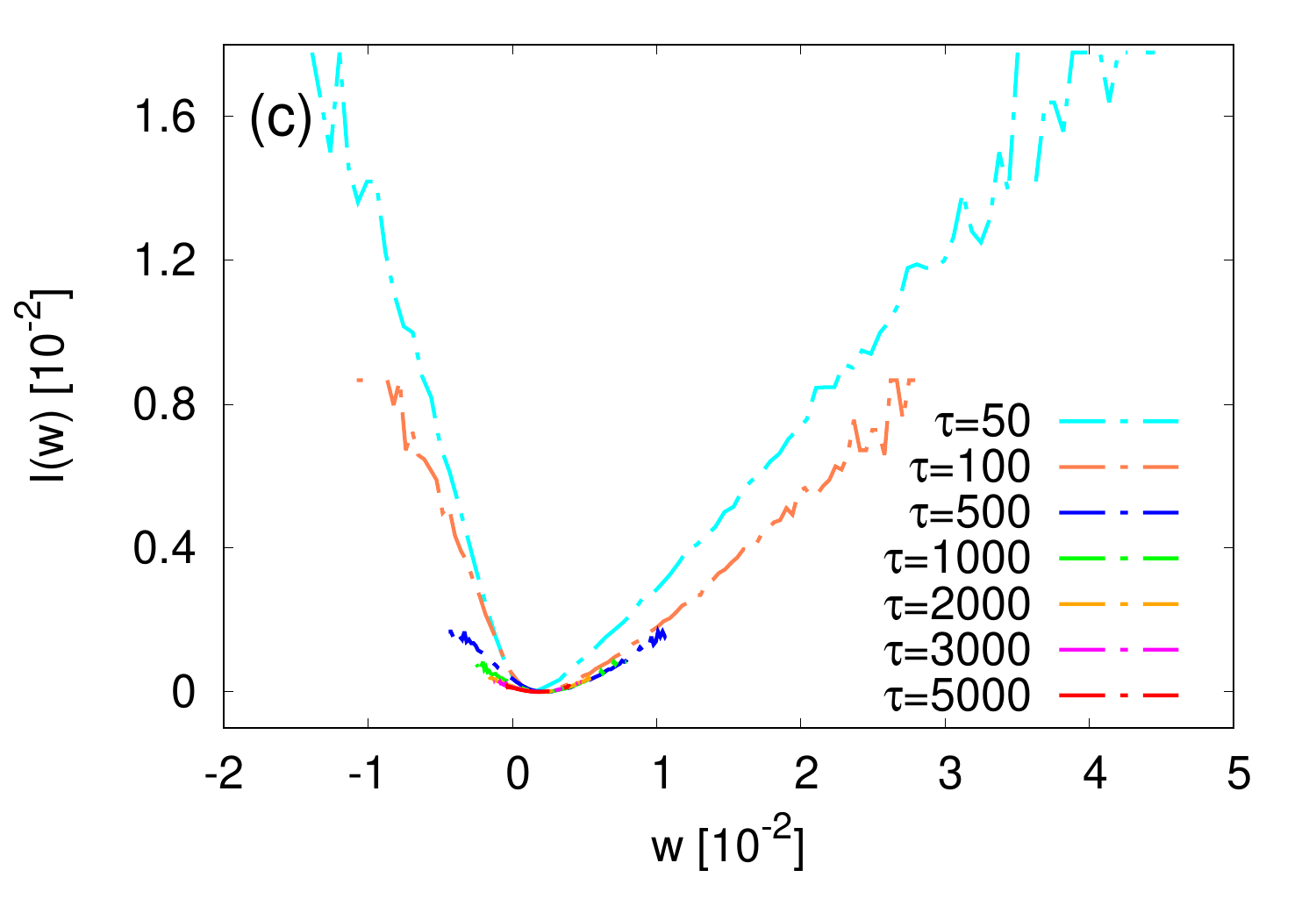}\qquad
       \includegraphics[width=0.45\columnwidth]{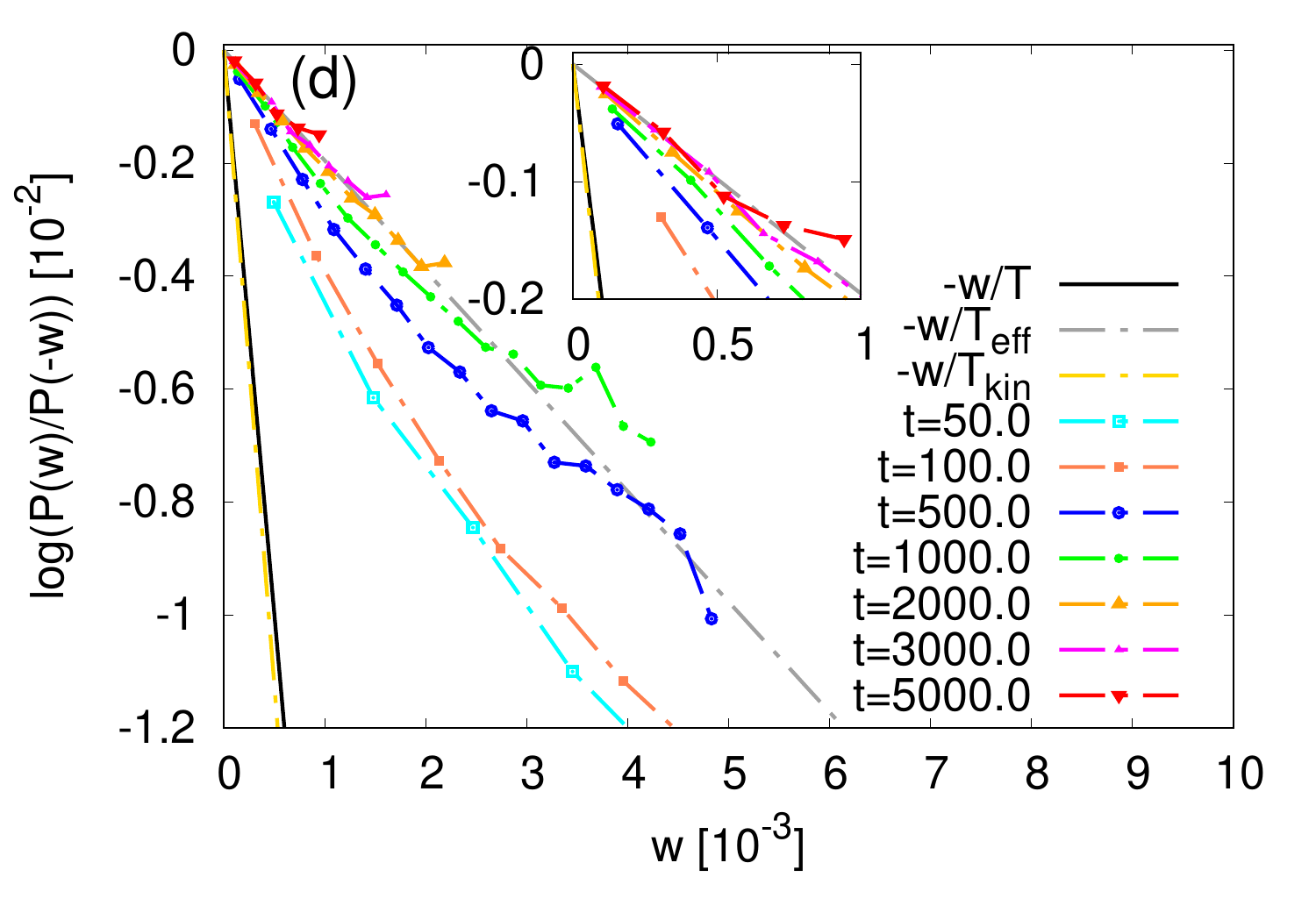}
  \end{tabular}
\caption{\footnotesize{$(a)$ and $(c)$: direct numerical estimates of the rate function of the active work for an AOUP confined by the anharmonic potential $k_{stiff}\bm{r}^{10}/10$ in $d=2$ respectively for the parameter choice $T=0.05$, $F_a=0.5$, $\gamma=10$, $k_{stiff}=1.0$, corresponding to a center-peaked position distribution, and $T=0.05$, $F_a=10.0$, $\gamma=100$, $k_{stiff}=1.0$, corresponding to a finite radius annular position distribution, at different simulation times, as reported in the legends. $(b)$ and $(d)$: plot of the rate function difference $I(w)-I(-w)$ corresponding respectively to the rate functions in $(a)$ and $(c)$. The black continuous line is $-w/T$, while the yellow dot-dashed line and the grey dot-dashed lines are respectively $-w/T_{eff}$ and  $-w/T_{kin}$ (see main text). The inset in $(d)$ shows an enlargement of the main figure around the origin. Note how in both cases the stationary form of the fluctuation theorem is reached after a few persistence times $\gamma_R^{-1}$.}}
\label{fig:fr-stiff}
\end{center}
\end{figure}

In this section we check the validity of the FR for the active work (per unit time) distributions in the free-particle case and in the presence of confining potentials. Note that the direct check of the FR through the rate function, using \autoref{DFTdirect}, requires the underlying numerical distribution to be significantly sampled for negative values. This condition is not satisfied for all the cases considered so far in this work, thus we resort to simulations with a reduced magnitude of the activity (parameters values are indicated in the caption of \autoref{fig:fr-free-harmonic} and \autoref{fig:fr-stiff}). Alternatively, we provide also an indirect check of the FR by using the SCGF data shown \autoref{fig:free-SCGF-rate-cloning} and \autoref{fig:harmonic-SCGF-cloning}, obtained with the previous choice of parameters trough the cloning algorithm (see \autoref{app1:cl_alg}), in \autoref{eq:scgf_symm}

%On the other hand, following a particle-based approach, \autoref{eq:active-work} represents also the amount of energy (heat) exchanged between the particle and the environment. Once divided for the environment temperature $T$, it can actually be considered as the total entropy production in absence of external potentials, or its active contribution in their presence \cite{eps_defs}.} 
 
In the free-particle case, a simple calculation shows that the SCGF \autoref{eq:SCGF} satisfies the symmetry property \autoref{eq:scgf_symm} with $c=-T^{-1}$, in fact
\begin{equation}
g\left(-\frac{1}{T}-\lambda\right)=\frac{d}{2}\left(\gamma_R-\sqrt{\gamma_R^2-4D'_R\gamma\left(1+\gamma D_T\lambda\right)\lambda}\right)=g(\lambda)~.
\end{equation}
Indeed, \autoref{fig:fr-free-harmonic}$(c)$ shows that the SCGF $g(\lambda)$ taken from \autoref{fig:free-SCGF-rate-cloning} and $g(-\lambda-T^{-1})$ overlap well, apart from the last points on the left-hand and on the right-hand side of the figure which are affected by the convergence problems detailed in \autoref{ssec:cloning}. Therefore, we can assert that a FR for the free-particle active work is satisfied with slope $c=-T^{-1}$. A direct proof of the validity of this theorem can also be obtained using the rate function \autoref{eq:rate-function}: after some algebra, we find the result $I(w)-I(-w)=-w/T$, valid for every choice of parameters. For the sake of completeness, panels $(a)$ and $(b)$ of \autoref{fig:fr-free-harmonic} provide numerical support to this result. In particular, panel $(a)$ shows the rate functions of the active work evaluated through direct sampling setting a lower activity, as indicated in the caption, while panel $(b)$ shows the FR at various times computed using rate functions from panel $(a)$.

The equivalence between the constant $c$ and the negative inverse bath temperature holds for a large class of models and dynamical observables~\cite{seifert_fl_th}. Among these observables there is the entropy production, which, let us remark, is proportional to the active work defined in \autoref{eq:active-work} provided the self-propulsion velocity $\bm{v}(t)$ is considered even under time reversal transformation \cite{dabelowAOUP, ketaLARGEDEV}. In general, one can define a fluctuation-relation temperature $T_{FR}$ such that $c=-T_{FR}^{-1}$, and compare it with other quantities such as the kinetic temperature $T_{kin}$, defined from the equipartition theorem, and the effective temperature $T_{eff}(t)$ defined from the fluctuation-dissipation theorem~\cite{let_temp, isa_teff1, isa_teff2}. Hereafter we consider the value of $T_{eff}(t)$ measured in the long time limit, where it reaches a constant value, so we drop the explicit time dependence (See \autoref{app2:teff_est} for more details). As reported in the appendix, we can compute these quantities analytically in the free AOUP case, and find that their values grow quadratically with the P\'eclet number with different functional forms, making them distinguishable with respect to each other and to the bath temperature. For the the free AOUP, we have that $T_{FR}=T$.

Concerning the confined AOUP, we resort to numerics to measure the $T_{FR}$ for some representative cases and compare it with the values of $T$, $T_{kin}$ and $T_{eff}$. For the harmonic potential, we checked directly the FR using a lower activity than that reported in~\autoref{fig:harmonic-SCGF-cloning} and~\autoref{fig:harmonic-rate-function}, in order to be able to sample also negative values of the active work (panel $(d)$ of \autoref{fig:fr-free-harmonic}) and evaluate \autoref{DFT}. From the results reported in panel $(e)$ it is evident that, similarly to the free-particle case, the slope of the curves is compatible with $T_{FR}=T$. We remark that in this low-activity case the bath temperature $T=0.05$ is essentially indistinguishable from the kinetic and effective temperatures $T_{kin}\simeq 0.0500$ and $T_{eff}\simeq 0.0505$, whose exact asymptotic expressions are reported in \autoref{app2:teff_est}. In the high-activity case we can estimate $T_{FR}$ by using the SCGF of~\autoref{fig:harmonic-SCGF-cloning}. Panel $(f)$ of \autoref{fig:fr-free-harmonic} shows that the SCGF satisfies the same symmetry property of the free-particle case $g(\lambda)=g(-\lambda-T^{-1})$ near the function minimum, so that $T_{FR}=T=0.05$. In the high-activity case the three temperatures are now perfectly distinguishable ($T=0.05$, $T_{kin}\simeq0.1804$ and $T_{eff}\simeq87.1199$), and thus we conclude that only the bath temperature is compatible with $T_{FR}$.

For the stiff potential $U_{stiff}(\r)=k_{stiff} \r^{10}/10$ we report in \autoref{fig:fr-stiff} the results of the analysis of the FR for two significative choices of parameters, giving rise respectively to a center-peaked stationary position distribution (low-activity) and to an annular one (high-activity), as described in \autoref{ssec:nonlinear}. In panels $(a)$ and $(b)$, the choice is such that the position distribution of the confined AOUP is peaked at the center, as in \autoref{fig:steady-state-stiff} center panel. In this case, as it can be seen from panel $(b)$, a FR is again satisfied with a slope compatible with $T_{FR}=T=0.05$, which is indistinguishable from $T_{kin}\simeq 0.0499(\pm0.01\%)$, but lower than $T_{eff}\simeq0.0717(\pm0.01\%)$ (both temperatures are measured numerically in this case). 
For higher activity (panels $(c)$ and $(d)$), the position distribution of the confined AOUP is annular with finite radius, as in \autoref{fig:steady-state-stiff} right panel. In this case, the FR reported in panel $(d)$ is again satisfied, but with a $T_{FR}\simeq 0.4887(\pm 4.9\%)$ much larger than the bath temperature $T=0.05$. Here, a comparison with $T_{kin}\simeq 0.0444(\pm0.001\%)$ and $T_{eff}\simeq 0.4775(\pm0.01\%)$ shows that $T_{FR}=T_{eff}$ within numerical accuracy. This difference between low and high activity can be justified by the fact that in the former case the particle dynamics takes place near the potential minimum, while in the latter case it does not.

\section{Conclusions}
\label{sec:conclusions}

In this paper we have studied the large deviations of the time-averaged power injected by the self-propulsion force---or active work---for an Active Ornstein-Uhlenbeck Particle (AOUP) interacting with a confining potential. We have examined, in particular, four cases: the free particle case, without any confining potential; the harmonic case, with a quadratic potential; two anharmonic potentials, namely a `stiff' potential growing as $\bm{r}^{10}$ and a circular well modelled with a WCA potential.

For the free-particle case, we have obtained the rate function of the active work in general dimension $d$ exactly from saddle-point calculations of the inverse Laplace transform of the CGF. 
%The generating function, in turn, is computed with path-integral techniques~\cite{pathintegral}. 
The corresponding SCGF was recently computed for $d\,{=}\,2$ in~\cite{grandpreENTROPY} by solving a tilted eigenvalue problem. The rate function of the free problem can also be obtained from the SCGF via Legendre-Fenchel transform. This might lead to erroneous results when the cumulant generating function displays diverging sub-exponential contributions~\cite{farago}. Our analysis takes such contributions into account, but we show that, at variance with the corresponding passive problem~\cite{farago}, the interplay between thermal and active noise causes the non-analyticities stemming from sub-exponential contributions to be covered by those of the SCGF, showing that estimating the rate function as a Legendre-Fenchel transform of the SCGF yields the correct result. 

For the confining potentials, we estimated the RF and the SCGF numerically. For a quadratic confining potential, we find the development of linear tails and a similar phenomenology of the free-particle case. These tails suggest the presence of a typical condensation transition due to a {\it sticking of the saddle-point} mechanism \cite{kac, condensation1, condensation2, condensation3, condensation4}, which would lead to a rate function growing linearly with the active work after a certain threshold---further analytical and numerical investigations will be needed to demonstrate this point.

For a general nonlinear confining potential, the AOUP will push against the slope of the potential in steady state when the P\'eclet number is sufficiently high. We characterized the SCGF of the active work for two representative nonlinear potentials and found that they do not show any singular behaviour despite accumulation effects at the boundaries. We found that lower-than-average fluctuations of the active work correspond to particles pushing against the potential and are realised with highly correlated kicks always pointing against the direction of the potential slope. Higher-than-average fluctuations correspond instead to particles which move in circles for most of the observation time. It would be of interest, in this respect, to extend the analysis to ensembles of interacting AOUPs. Such systems are known to display similar macroscopic average properties to those of system of interacting Active Brownian Particles~\cite{ketaLARGEDEV}, thus it is a natural question whether the analogy extends to features related to rare fluctuations~\cite{gonn1, nemotoLARGEDEV}.

%\textcolor{red}{Per la prima volta le fluttauzioni del lavoro sono state studiate in questo caso, scrivere in modo chiaro dando importanza a questa cosa}
%For a general nonlinear confining potential, when the P\'eclet number is sufficiently high, the AOUP will be pushing against the slope of the potential in steady state. As a result, new mechanisms for active work fluctuations appear which overshadow those which were dominant in the free and harmonic problem. In the latter problems, for instance, lower-than-average fluctuations are realised by having a sequence of anti-correlated kicks from the active noise, whereas in the other cases, where the AOUP is mostly found pushing against the potential's slope, lower-than-average fluctuations can be realised with highly correlated kicks, so that the particle remains stuck against the potential's slope. Similarly, fluctuations with large active work are realised with highly correlated kicks in the free and harmonic problems and with anti-correlated kicks in a general anharmonic potential. This is a major difference with respect to the corresponding passive problem~\cite{farago}, where both the functional form of the rate function and the phenomenology of fluctuations are independent of the confining potential. Despite the different phenomenology, no special structures appear in the rate functions \textcolor{blue}{that are captured by our numerical techniques}
%, or at least none that our numerical techniques can capture. 

Finally, we have studied the fluctuation relation for active work fluctuations in three different situations: the free-particle case, the harmonically confined AOUP and the AOUP confined by the stiff anharmonic potential. In particular, we defined a typical temperature $T_{FR}$ stemming from the fluctuation relation slope.
For the free-particle case, we proved analytically that the fluctuation relation is satisfied with $T_{FR}=T$, and supported the results with numerical simulations. For the harmonically confined AOUP, we found numerically that the fluctuation relation is satisfied with $T_{FR}=T$, and that this result is independent of the activity.
For the anharmonically confined AOUP, we found that low-activity AOUP satisfies the fluctuation relation with $T_{FR}=T$, while high-activity AOUP have a much higher value of $T_{FR}$ than the bath temperature, which is equal to an effective temperature $T_{eff}$ estimated independently using the fluctuation-dissipation theorem. This increase in the effective temperature corresponds to giving rise to an annular position distribution instead of a center-peaked one when increasing the activity.

These results could be connected in some way to recent works showing that in the AOUP model the entropy production rate is different from zero only when the particle is under the action of a more-than-quadratical potential \cite{capriniAOUP, martinAOUP, fodorHOWFAR}. In order to better understand this possible connection, the role of the effective temperature and the general behaviour of active systems, a more in-depth analysis of fluctuation theorems for active systems could be of interest. We then leave as a future possible work further extensions of our results to analytical and numerical checks of fluctuation relations.

\appendix

\section{Details on numerical integration methods}
\label{app1:cl_alg}
The first method we applied consists in performing numerical integration in two dimensions of the equations
\begin{equation}\label{eq:AOUP1-overd}
 m\ddot{\bm{r}}(t)=-\gamma\dot{\bm{r}}(t) + \gamma\bm{v}(t) -\nabla U(\bm{r}(t), t)+ \gamma\sqrt{2 D_T}~\bm{\xi}(t),
\end{equation}
which includes the inertial term $m\ddot{\bm{r}}(t)$, with $m$ the mass of the particle and in which different potentials expressions are considered. The equations are integrated using the Vanden-Eijnden- Ciccotti algorithm~\cite{ciccotti}. The parameter choice is different, as indicated in the figures caption in each section, but in general we choose the mass $m=1$ and the friction coefficient $\gamma$ in such a way that the ratio $m/\gamma<<1$ and the system can always be considered as effectively overdamped for timescales larger than the inertial time $t_I\,{=}\,m/\gamma\,{=}\,0.1$. We use as integrating timestep $dt\,{=}\,0.01$, and evolve the system up to a maximum observation time of order $\tau\,\sim 10^3$, or $10^5 \, dt$, computing the active work directly from the definition~\autoref{eq:active-work}. The empirical distributions of the active work are obtained by integrating the dynamics for $N_c\,{=}\,10^5$ independent realisations of the stochastic forces, with initial conditions sampled from equilibrated systems evolved for $\tau_{eq}\,{=}\,10^2>>t_I$. 

The second method we use to estimate the rate function is the cloning algorithm, an algorithm based on importance  sampling  ideas and described briefly as follows. A large number $N_c$ of copies, or clones, of the system is evolved simultaneously using the same integration scheme described in the previous paragraph. Copies are then cloned or pruned depending on the value of the active work so as to generate a biased ensemble with a lower- or higher-than-usual average active work. Specifically, we divide the entire simulation time interval $\tau$ in $M$ subintervals such that $\tau=M\, \Delta t$, with $\Delta t=1$ (corresponding to 100 timesteps).  We compute the active work of each copy after every time interval $\Delta t$ and we register the values $\Delta W_t^a\,{=}\,W^a_{t+\Delta t}-W^a_t$,  with $a$ the copy index, together with the weighted sum over all copies $G_t(\lambda)\,{=}\,\sum_b e^{\lambda \, \Delta W^b_t}$. $\lambda$ is the biasing parameter, coinciding with the argument of the cumulant generating function. Thus an integer score $n^a$ is assigned to each copy according to
\begin{equation}
 n_a=\left\lfloor \frac{e^{\lambda \Delta W^a_t}}{G_t(\lambda)} N_c+\zeta\right\rfloor~,
\end{equation}
where $\zeta$ is a random number uniformly distributed in $[0,1]$ and $\lfloor.\rfloor$ denotes the lower integer part. For $\lambda$ positive, trajectories with a higher $\Delta W^a_t$ will have a higher score. Each copy $a$ with positive score is then cloned so that it appears $n_a$ times, whereas copies with $n_a\,{=}\,0$ are pruned. After this step, copies are deleted or cloned at random in order to keep the number of copies constant. After repeating this evolve-and-clone procedure for the $M$ steps, the SCGF can be estimated as
\begin{equation}\label{eq:cloning-SCGF}
	g_\tau(\lambda) = \frac{1}{\tau}\log\hat\Pi(\lambda) \simeq \frac{1}{(M-1) \Delta t} \sum_{t=0}^{M-1} \log\left(\frac{G_t(\lambda)}{N_c}\right).
\end{equation}

\section{Effective and Kinetic temperature}
\label{app2:teff_est}
In this appendix we recall how the effective and kinetic temperatures can be defined on the basis of the fluctuation-dissipation relation and of the equipartition theorem (see e.g. \cite{let_temp, isa_teff1, isa_teff2}). Concerning the effective temperature, the starting point is just the equilibrium fluctuation-dissipation relation 
\begin{equation}
2T\chi(t',t)=\Delta^2(t',t)~,
\label{eq:fdt_eq}
\end{equation}
where
\begin{equation*}
\Delta^2(t',t)=\braket{[\r(t)-\r(t')]^2}
\end{equation*}
is the total mean square displacement and
\begin{equation*}
\chi(t',t)=\int_{t'}^t dt'' \sum_{\alpha=1}^dR_{\alpha\alpha}(t'',t)
\end{equation*}
is the integrated linear response, with
\begin{equation*}
R_{\alpha\beta}(t',t)=\frac{\delta\braket{r_\alpha(t)}_h^\lambda}{\delta h_\beta^\lambda(t')}\bigg|_{h_\beta^\lambda=0}
\end{equation*}
linear response of the system, $\alpha, \beta$ dimension indices, $d$ dimension of the system and $h_\beta^\lambda(t')$ an external perturbation depending on the parameter  $\lambda$. The idea is to exploit \autoref{eq:fdt_eq} to define in the active out-of-equilibrium system the time dependent effective temperature 
\begin{equation}
T_{eff}(t',t)=\frac{\Delta^2(t',t)}{2\chi(t',t)}~.
\label{eq:teff_def}
\end{equation}
The kinetic temperature is instead defined exploiting the velocity fluctuations and the equipartition theorem. From the latter, we can in fact write for each degree of freedom $i$ of a system that
\begin{equation*}
    \frac{1}{2}m \braket{\dot{\r}_i^2(t)}=\frac{1}{2}k_BT_{kin}(t)~,
\end{equation*}
where $k_B$ is the Boltzmann constant and $\braket{\ldots}$ denotes an ensemble mean. From this expression one simply obtains the kinetic temperature definition as
\begin{equation}
    T_{kin}(t)=\frac{m \braket{\dot{\r}_i^2(t)}}{k_B}~.
    \label{eq:tkin_def}
\end{equation}

In the free-particle case, we can use the expressions for the mean square displacement and the mean value of the squared velocity reported in \cite{inertial_aoup}, and the computed integrated response function 
\begin{equation*}
\chi(t',t)=\frac{t-t'}{\gamma}~,
\end{equation*}
with $\gamma$ the bath friction coefficient. We find that in the limit $t\rightarrow\infty$
\begin{equation*}
    T_{kin}(t) \quad \longrightarrow\quad T+\left(\frac{F_a}{m}\right)^2\frac{D'_R}{\gamma_R}\frac{1}{\gamma}\frac{1}{(\gamma_R+\frac{\gamma}{m})}
\end{equation*}
and
\begin{equation*}
    T_{eff}(t) \quad \longrightarrow\quad T+\left(\frac{F_a}{m}\right)^2\frac{D'_R}{\gamma_R}\frac{1}{\gamma^2\gamma_R}~,
\end{equation*}
with $t'=0$. In these formulas both $T_{kin}(t)$ and $T_{eff}(t)$ reach a constant value for large enough times and have both a quadratic dependence on the $Pe$ but with a different functional form, such that they can be distinguished with respect to each other and from the bath temperature $T$ when Pe is sufficiently high.
%As the activity is increased, due to their functional forms and the quadratic dependence on the P\'eclet number, such temperatures become distinguishable with respect to each other and from the bath temperature $T$, the latter being, as proved in \autoref{sec:fl_th}, the only one satisfying the fluctuation relation in the free-particle case for every possible choice of parameters.}

We can provide the expression for $T_{eff}(t)$ and $T_{kin}(t)$ also in the harmonically confined case ($U(\r)=k \r^2/2$). Using the expressions for the mean square displacement and the mean value of the squared velocity reported in \cite{inertial_aoup} for a harmonically confined AOUP, and the integrated linear response
\begin{equation*}
\chi(t',t)=\frac{1-e^{-\frac{\gamma}{k}(t'-t)}}{k}~,
\end{equation*}
with $\gamma$ the bath friction coefficient and $k$ the elastic constant of the harmonic potential, we find that in the limit $t\rightarrow\infty$ 
\begin{equation*}
    T_{kin}(t) \quad \longrightarrow\quad T+\left(\frac{F_a}{m}\right)^2\frac{D'_R}{\gamma_R}\frac{1}{\gamma}\frac{1}{(\gamma_R+\frac{\gamma}{m}+\frac{k}{m \gamma_R})}
\end{equation*}
and
\begin{equation*}
    T_{eff}(t) \quad \longrightarrow\quad T+\left(\frac{F_a}{m}\right)^2\frac{D'_R}{\gamma_R}\frac{1}{\gamma}\frac{1+\frac{\gamma}{m\gamma_R}}{(\gamma_R+\frac{\gamma}{m}+\frac{k}{m \gamma_R})}~,
\end{equation*}
with $t'=0$. Similarly to the free-particle case, both temperatures reach a constant value for large times.

\begin{figure}[h!]
\begin{center}
  \begin{tabular}{cc}
       \includegraphics[width=0.5\columnwidth]{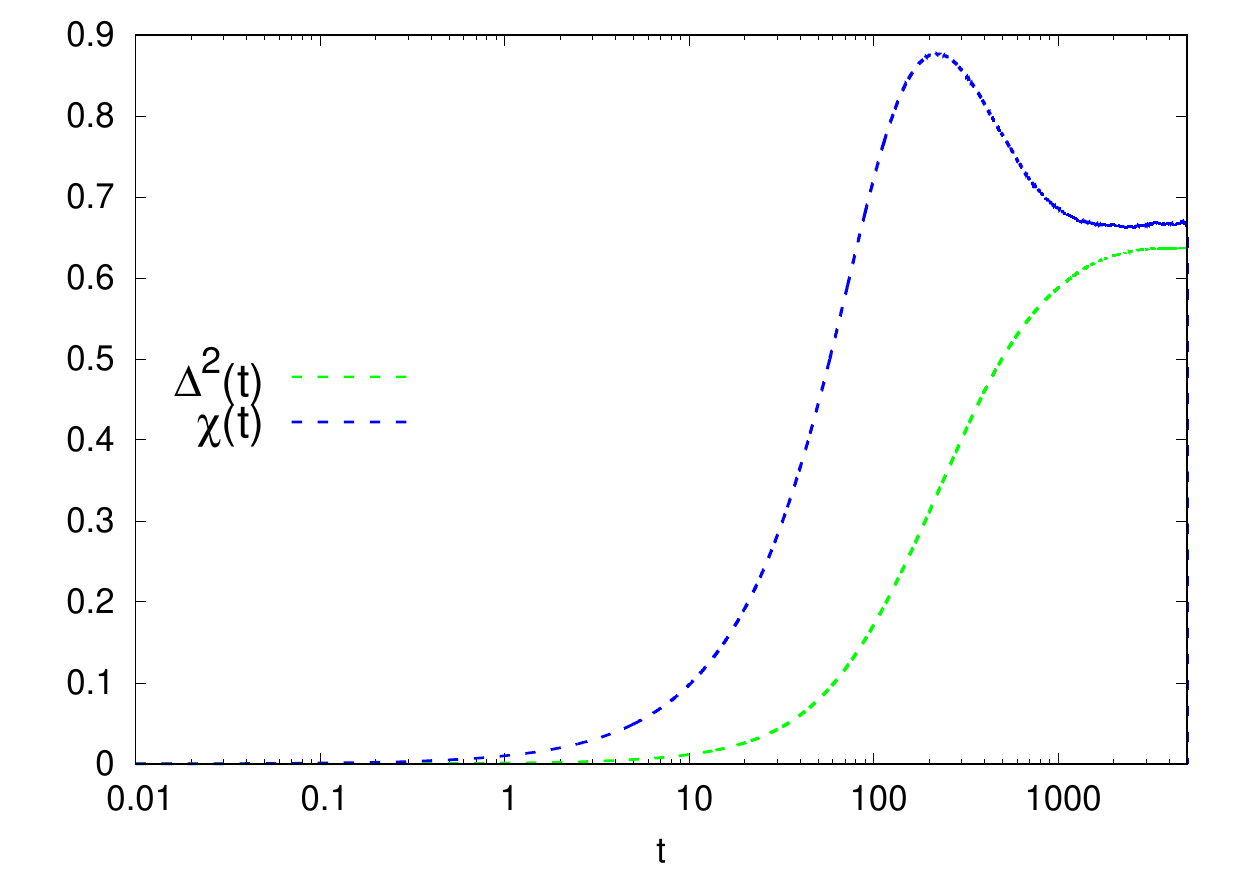}
       \includegraphics[width=0.5\columnwidth]{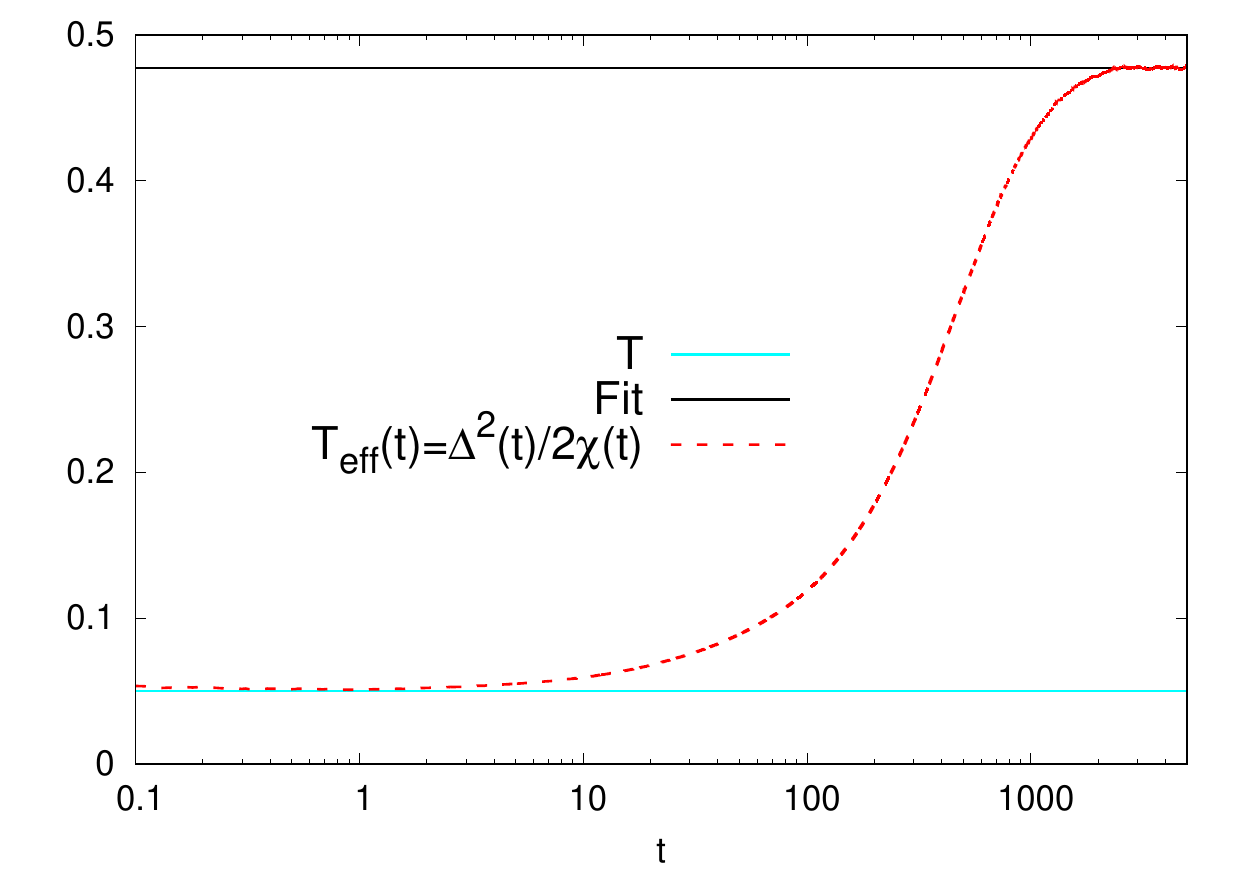}\\
  \end{tabular}
\caption{\footnotesize{Left: direct numerical estimates of the $x$-component mean square displacement $\Delta^2(t)$ (green dashed line) and of the $x$-component integrated linear response $\chi(t)$ (blue dashed line) for an AOUP in the stiff potential ($U_{stiff}(\r)=k_{stiff}\r^{10}/10$) for the parameter choice $T=0.05$, $F_a=10.0$, $\gamma=100$, $k_{stiff}=1.0$. The integrated linear response is obtained simulating the AOUP in presence of a constant force $h=0.2$ acting in the $x$ direction and considering $N_c=3\cdot 10^6$ independent evolutions of the biased system. Right: effective temperature $T_{eff}(t)$ (red dashed line) obtained from \autoref{eq:teff_def}, using $\Delta^2(t)$ and $\chi(t)$ of the left panel. As a reference, we report also the value of the bath temperature $T$ (cyan solid line) and the fitted effective temperature at large times $T_{eff}=0.4775(\pm0.01\%)$ (black solid line).}}
\label{fig:teff_stiff}
\end{center}
\end{figure}

In the anharmonically confined case ($U_{stiff}(\r)=k_{stiff}\r^{10}/10$), we estimated $T_{eff}(t)$ and $T_{kin}(t)$ numerically. The former was estimated directly using the definition \autoref{eq:tkin_def}.  The latter was estimated from the definition~\autoref{eq:teff_def}, measuring independently $\Delta^2(t)$ and $\chi(t)$. $\Delta^2(t)$ was measured as the x-component mean square displacement, while for $\chi(t)$ we applied a constant force $h$ in the $x$-direction. We chose $h=0.2$ in order for the force to be small enough to remain in the linear regime and high enough to overcome the large fluctuation effects. In \autoref{fig:teff_stiff} left panel we report $\chi(t)$ and $\Delta^2(t)$ in a high-activity case with  $F_a=10.0,~\gamma=100,~T=0.05$. Note that these functions reach a constant value after a few permanence times $\gamma_R^{-1}$.  In \autoref{fig:teff_stiff} right panel we report instead $T_{eff}(t)$. Notice that for times smaller than the persistence time, $t<<\gamma_R^{-1}$,  $T_{eff}(t)=T$, with $T$  the bath temperature.  For $t>>\gamma_R^{-1}$, instead, we reach a fitted constant value of $T_{eff}(t)\simeq 0.4775(\pm0.01\%)$. $T_{kin}$ is instead constant over time and has a fitted value of $T_{kin}\simeq0.0444(\pm0.001\%)$. We also estimated the two temperatures for a low-activity case, with $F_a=0.5,~\gamma=10,~T=0.05$. In this case the numerical measurements yield $T_{kin}\simeq0.0499(\pm0.01\%)$ and for large times $T_{eff}\simeq 0.0717(\pm0.01\%)$.

%\section*{References}
\newpage
\bibliographystyle{iopart-num}		% The reference style
\bibliography{Tesi.bib} 	        % Multiple bib files

\end{document}